\documentstyle[preprint,aps,psfig]{revtex} 
\tighten 
\draft 
\begin{document} 
\preprint{\today} 
\title{ Test of
mode coupling theory for a supercooled liquid of diatomic molecules. II.
$q$-dependent orientational correlators } 
\author{Stefan K\"ammerer, Walter Kob and Rolf Schilling} 
\address{Institut f\"ur Physik, Johannes Gutenberg-Universit\"at, 
		Staudinger Weg 7, D-55099 Mainz, Germany}
\maketitle

\begin{abstract}
Using molecular dynamics computer simulations we study the dynamics of
a molecular liquid by means of a general class of time-dependent
correlators $S_{ll'}^m(q,t)$ which explicitly involve translational
(TDOF) and orientational degrees of freedom (ODOF). The system is
composed of rigid, linear molecules with Lennard-Jones interactions.
The $q$-dependence of the static correlators $S_{ll'}^m(q)$ strongly
depend on $l$, $l'$ and $m$. The time dependent correlators are
calculated for $l=l'$. A thorough test of the predictions of mode
coupling theory (MCT) is performed for $S_{ll}^m(q,t)$ and its self
part $S_{ll}^{(s)m}(q,t)$, for $l=1,\ldots,6$. We find a clear
signature for the existence of a single temperature $T_c$, at which the
dynamics changes significantly.  The first scaling law of MCT, which
involves the critical correlator $G(t)$, holds for $l \ge 2$, 
but no critical law is observed.  Since this is true
for the same exponent parameter $\lambda$ as obtained for the TDOF, we
obtain a consistent description of both, the TDOF and ODOF, with the
exception of $l=1$. This different behavior for $l \ne 1$ and $l=1$
can also be seen from the corresponding susceptibilities
$(\chi'')_{ll}^m(q,\omega)$ which exhibit a minimum at about the same
frequency $\omega_{min}$ for all $q$ and all $l \ne 1$, in contrast to
$(\chi'')_{11}^m(q,\omega)$ for which $\omega'_{min} \approx 10 \;
\omega_{min}$ . The asymptotic regime, for which the first scaling law
holds, shrinks with increasing  $l$. The second scaling law of MCT
(time-temperature superposition principle) is reasonably fulfilled for
$l \ne 1$ but not for $l = 1$. Furthermore we show that the $q$- and
$(l,m)$-dependence of the self part approximately factorizes, i.e.
$S_{ll}^{(s)m}(q,t) \cong C_l^{(s)}(t) \; F_s(q,t)$ for all $m$.

\end{abstract}

\narrowtext \pacs{PACS numbers: 61.43.Fs, 61.20.Ja, 02.70.Ns, 64.70.Pf}

\section{Introduction} 
\label{sec1}

In the last few years quite a few papers were published in which
computer simulations were used to study the time dependence of the {\it
translational} degrees of freedom (TDOF) in supercooled liquids.  On
the other hand, the {\it orientational} degrees of freedom (ODOF) were
so far investigated in much less detail since the simulation and data
analysis of systems in which the particles are molecules are quite a
bit more involved than the ones in which the particles have no
structure.  However, since most real materials are of molecular nature
and since experimental methods such as light scattering or dielectric
measurements probe also the ODOF, it is important to understand how the
dynamics of the TDOF and the ODOF are related to each other. Only by
understanding this relationship it will be possible to make a correct
interpretation of the experimental measurements and to gain insight
into the nature of the glass transition, i.e. the dramatic slowing down
of the relaxation dynamics of supercooled liquids upon approaching the
glass transition temperature. A more thorough discussion of these
connections can be found, e.g., in Ref.~\cite{r1}, were we also review
some of the other work in this field.

Very recently we have carried out a molecular dynamics computer
simulation of a simple molecular system in order to make a detailed
comparison between the dynamics of the TDOF and the
ODOF~\cite{r1}. Each molecule in this system is dumb-bell
shaped and consists of two Lennard-Jones particles that are separated
by a fixed distance $d$.  More details on the system and the simulation
can be found in Ref.~\cite{r1}. In that paper we studied the
time and temperature dependence of the orientational correlation
functions

\begin{equation}
C_l(t) = \frac{1}{N} \sum_{n,n'} \langle
                P_{l}(\vec{u}_n(0) \cdot \vec{u}_{n'}(t)) \rangle
                \qquad , \qquad l\geq 1 \qquad,
\label{eq2}
\end{equation}
and the self part $C_l^{(s)}(t)$. Here $\vec{u}_n(t)$ is the unit
vector pointing along the molecular symmetry axis of molecule $n$ and
$P_l(x)$ is the $l$-th Legendre polynomial. The relevance of these type
of correlation functions is given by the fact that they can be measured
in experiments. The main results of that paper were that the
temperature dependence of the relaxation times of $C_l$, $C_l^{(s)}$
and the diffusion constant $D$ were given by a power law with the same
critical temperature $T_c$ but with critical exponents that
depend on the observable. In addition we showed that the so-called
time temperature superposition principle works well for $C_l^{(s)}$, if
$l > 2$. Thus we concluded that many of the predictions of mode-coupling
theory (MCT)~\cite{mct,gotze91} hold for these correlation functions,
although certain discrepancies are present.

In the preceding paper, subsequently called KKSI, we have investigated
the time and temperature dependence of the {\it translational} degrees
of freedom by studying quantities like the van Hove correlation
function $G(r,t)$ and the intermediate scattering function
$F(q,t)$~\cite{r2}. The main conclusion of that paper was
that MCT is able to give also a good description for the time and
temperature dependence of these correlation functions. 

As we will demonstrate below, the intermediate scattering function
$F(q,t)$ and the orientational correlation functions $C_l(t)$ are just
a special case of a more general type of correlation function, which
involves the translational as well as the orientational degrees of
freedom at finite wave-vector $\vec{q}$, i.e.  
$\left| \vec{q} \right| > 0$. The goal of the {\it
present} paper is therefore to investigate the time and temperature
dependence of these more general correlation functions, since it is
these correlators which are needed for a more detailed description of the
dynamics of a molecular system. In addition these correlation functions
can also be calculated directly within the framework of MCT (although
such a calculation might be in practice quite involved) thus allowing
to perform a more stringent test of whether MCT is able to give a
correct description of the dynamics of the system investigated.

Our paper is organized as follows: In the next section we will
introduce the mentioned generalized correlation functions and will
discuss some of their properties. Section~\ref{sec3} presents the
results and the MCT-analysis and the final section contains a summary
and our main conclusions.

\section{correlation functions}
\label{sec2}

We introduce a set of correlators which involves the one-particle
density (including the angular dependence) for a molecular liquid of 
rigid, axially symmetric molecules:
\begin{equation}
\rho(\vec{x},\Omega,t) = \sum_{n=1}^{N} \delta(\vec{x}-\vec{x}_n(t))
\, \delta(\Omega,\Omega_n(t))
\label{eq5}
\end{equation}
where $\vec{x}_n(t)$ and $\Omega_n(t) \equiv (\theta_n(t),\phi_n(t))$ 
denote the center of mass position and the orientation of the 
$n$-th molecule at time $t$, respectively. Due to the non-Euclidean metric 
for the angles $\theta$ and $\phi$, one must use the invariant delta 
function $\delta(\Omega,\Omega')$. For this
and other details of the theoretical description of molecular liquids,
the reader is referred to the textbook by Gray and Gubbins \cite{r4}.
Expansion of $\rho(\vec{x},\Omega,t)$ with respect to a product of 
plane waves and spherical harmonics $Y_{lm}(\Omega)$ leads to the 
tensorial density modes
\begin{equation}
\rho_{lm}(\vec{q},t) = i^l \sqrt{4\pi}  \sum_{n=1}^{N}
e^{i \vec{q} \cdot \vec{x}_n(t)}  \; Y_{lm}(\Omega_n(t)) \qquad ,
\label{eq6}
\end{equation}
where $l=0,1,2,\ldots$ and $-l \leq m \leq l$.
The factor $\sqrt{4\pi}$ is used so that $\rho_{00}(\vec{q},t)$ 
equals the definition of
$\rho(\vec{q},t)$ for simple liquids and $i^l$ is introduced 
for convenience (see below). The corresponding correlators
\begin{equation}
S_{lm,l'm'}(\vec{q},t) = \frac{1}{N} \langle
\delta \rho^{\ast}_{lm}(\vec{q},t)
\: \delta \rho_{l'm'}(\vec{q},0)  \rangle
\label{eq7}
\end{equation}
of the fluctuation $\delta \rho_{lm}(\vec{q},t) = 
\rho_{lm}(\vec{q},t) - \langle \rho_{lm}(\vec{q},t) \rangle$ vanish 
for $(q,l,m) = (0,0,0)$, and are otherwise given by:
\begin{equation}
S_{lm,l'm'}(\vec{q},t) = \frac{4 \pi}{N}
	\; i^{l'-l} \sum_{n,n'} \left< \exp \left[-i \vec{q} \cdot 
	( \vec{x}_n(t) - \vec{x}_{n'}(0) ) \right]
	\, Y_{lm}^{\ast}(\Omega_n(t) ) \, Y_{l'm'}(\Omega_{n'}(0) )
	\right>
\label{eq8}
\end{equation}
which shows the explicit dependence on both, the TDOF and the ODOF. Its
corresponding self part $S_{lm,l'm'}^{(s)}(\vec{q},t)$ is obvious.

Taking into account that $Y_{00}=1/\sqrt{4\pi}$ one obtains from 
Eq.~(\ref{eq8}):
\begin{equation}
\frac{S_{00,00}(\vec{q},t)}{S_{00,00}(\vec{q})} = F(q,t) \quad ,
\label{eq9}
\end{equation}
i.e. the normalized density correlator for the center of mass
positions, which was studied in KKSI. On the other hand, we find from
Eq.~(\ref{eq8}) for $\vec{q}=0$:
\begin{equation}
S_{lm,l'm'}(0,t) = C_l(t) \delta_{mm'} \delta_{ll'} \quad .
\label{eq10}
\end{equation}
Here the addition theorem for the spherical harmonics \cite{r4} and the
isotropy have been used. As already mentioned in the Introduction, this
special case was investigated in Ref.~\cite{r1}. Eqs.~(\ref{eq9}) and
(\ref{eq10}) hold for the corresponding self part, as well.

Although it is not obvious how these correlators for $l,l' \neq 0$ can
be measured in real experiments for $\vec{q}\neq 0$, they are the basic
quantities which enter the MCT for a molecule in a simple liquid
\cite{r5} and for molecular liquids \cite{r6,r7,r8,kk,r9}.  To our
knowledge, there exists only one computer simulation which considers
$q$-dependent orientational correlators \cite{r10}. But the
experimental relevance of these correlators considered in
Ref.~\cite{r10} is unclear. The correlators given in Eq.~(\ref{eq8})
simplify a bit, if one uses the $q$-frame \cite{r4}, i.e.
$\vec{q}=\vec{q_0}\equiv(0,0,q)$. In that case one obtains \cite{r7}:
\begin{equation}
S_{lm,l'm'}(\vec{q}_0,t) \equiv S_{ll'}^{m}(q,t) \; \delta_{mm'} \quad ,
\label{eq11}
\end{equation}
which differ from zero only for $0 \leq \left|m\right| \leq
\mbox{min}(l,l')$ . Since $S_{ll'}^{m}(q,t) = S_{ll'}^{-m}(q,t)$, one
can restrict oneself to  $m \geq 0$. The introduction of $i^l$ in
Eq.~(\ref{eq6}) makes $S_{ll'}^{m}(q,t)$ a real quantity. The same
properties hold for the self part as well. In the following we will
present all results in the $q$-frame. 

Some of the equations that we will subsequently make use of have been
given in KKSI and are not reproduced here. We will refer to the $n$th
equation of that paper by (I-$n$).

\section{results}
\label{sec3}

This section is subdivided into two parts. The first part contains the
results for the static correlators $S_{ll'}^{m}(q)$, 
and the second one presents the dynamical correlators $S_{ll'}^{m}(q,t)$ 
and $S_{ll'}^{(s)m}(q,t)$. In the following we restrict the values of
$l$ and $l'$ to 0, 1 and 2.

\subsection{Static properties}

\begin{figure}[f]
\psfig{file=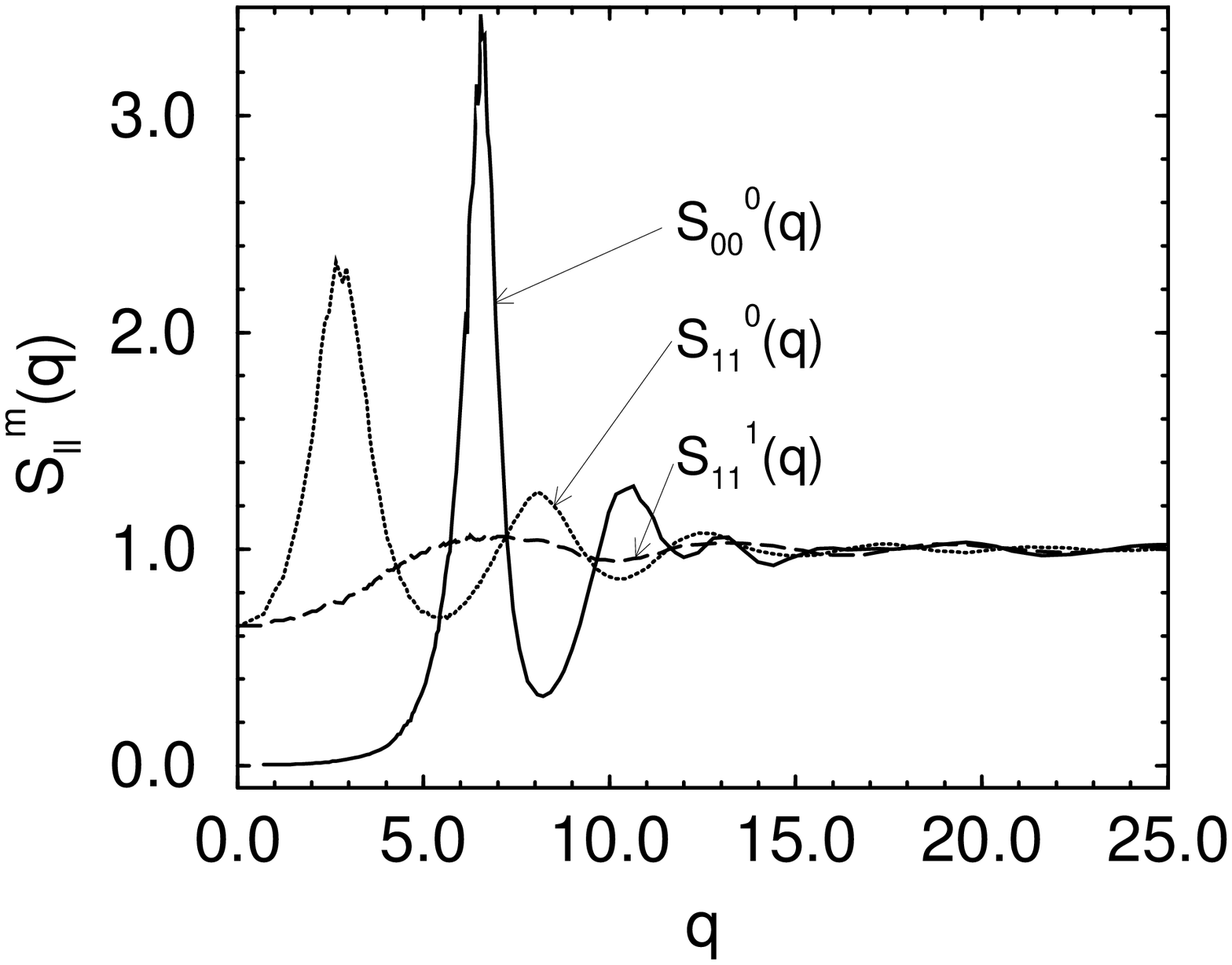,width=13cm,height=9cm}
\caption{
Wave vector dependence of static correlation functions for $T=0.477$.
 $S(q) \equiv S_{00}^0(q)$ (solid line),
$S_{11}^m(q)$ for $m=0$ (dotted line) and $m=1$ (dashed line).}
\label{fig2}
\end{figure}

\begin{figure}[f]
\psfig{file=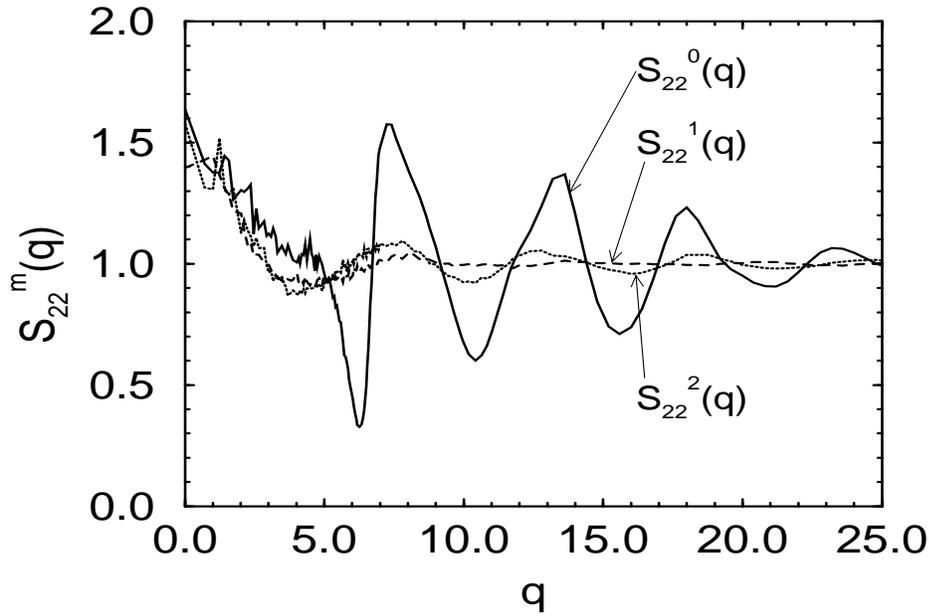,width=13cm,height=9cm}
\caption{
$S_{22}^m(q)$ versus $q$ for $T=0.477$ and $m=0$ (solid line),
$m=1$ (dashed line) and $m=2$ (dotted line).}
\label{fig3}
\end{figure}

\begin{figure}[f]
\psfig{file=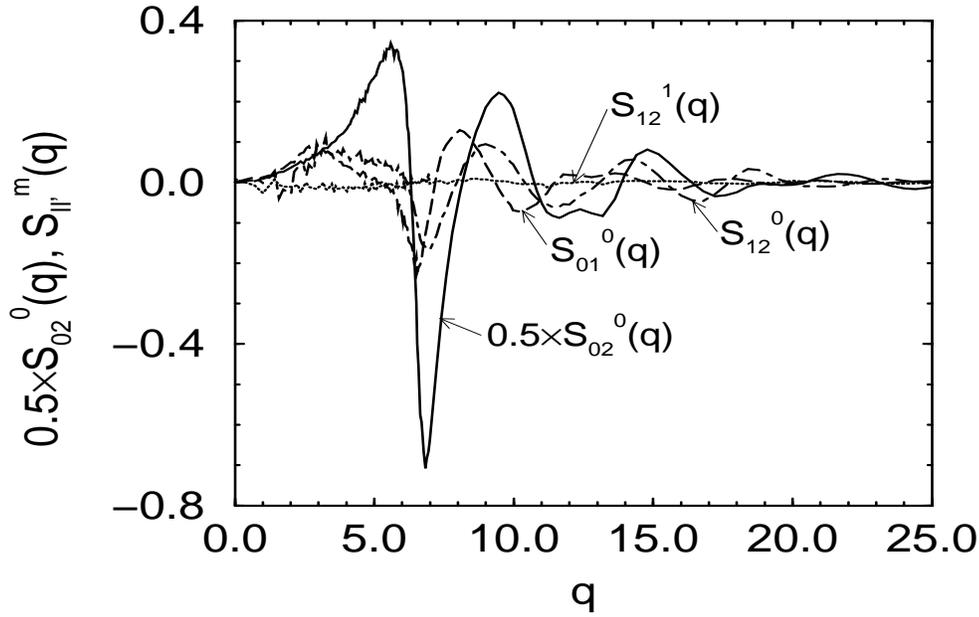,width=13cm,height=9cm}
\caption{
$0.5 \; \cdot \; S_{02}^0(q)$ (solid line),
$S_{01}^0(q)$ (dashed line), $S_{12}^0(q)$ (dashed dotted line) and
 $S_{12}^1(q)$ (dotted line) versus $q$ for $T=0.477$.}
\label{fig4}
\end{figure}

The static correlators are shown in Figs.~\ref{fig2} -~\ref{fig4} for
the lowest investigated temperature $T = 0.477$. First of all, it becomes obvious
from these figures that $S_{ll'}^{m}(q=0)$ is $m$-independent and
diagonal in $l$ and $l'$, as it should be due to isotropy. A comparison
of the various diagonal correlators in Figs.~\ref{fig2} and \ref{fig3}
with each other shows, that the correlators $S_{ll}^{0}(q)$ for $l=1$
and 2 possess a significant $q$-dependence similar to that of 
$S(q) \equiv F(q,0)$,
in contrast to those for $m \neq 0$. The same behavior was found for a
system of dipolar hard spheres~\cite{r7}, although for that system the
most prominent peak occurs for $S_{ll}^{1}(q)$ at $q=0$.  In contrast
to $S(q)$ and $S_{11}^{0}(q)$, the correlator $S_{22}^{0}(q)$ has a
rather broad maximum at $q=0$ with a height which is comparable to that
at $q'_{max}\approx 7.3$, the location of the main peak in
$S_{22}^0(q)$.  In Fig.~\ref{fig4} we present the non-diagonal
correlators $S_{ll'}^{m}(q)$ with $l \neq l'$. First of all one
recognizes that $S_{02}^{0}(q)$ is much larger than $S_{01}^{0}(q)$ and
$S_{12}^{m}(q)$. This can easily be understood. If the molecules had
``head-tail''-symmetry, then it can be shown that $S_{ll'}^{m}(q)
\equiv 0$, for $l,l'$ such that $l+l'$ is odd. Since for our molecules
this symmetry is only slightly broken, we expect $S_{ll'}^{m}(q)$ to be
much smaller for $l+l'$  odd than for $l+l'$ even.

The second point one recognizes from this figure is that the
non-diagonal correlators $S_{ll'}^{m}(q)$ can have the same magnitude
than the diagonal ones. Hence, there is no reason why the former should
be neglected in analytical calculations.  For example, since the
solutions of the MCT-equations for the time-dependent correlators
$S_{ll'}^{m}(q,t)$ are determined by the static correlators
$S_{ll'}^{m}(q)$, it might not be a good approximation to consider $l=l'$,
only.

\subsection{Dynamical properties}

We have investigated both, the self correlators for $l=l'=0,1,\ldots,6$ and 
the collective correlators for $l=l'=0,1$ and 2. Let us start with the self part 
$S_{ll}^{(s)m}(q,t)$. Often it is assumed (see e.g. \cite{r11}) that the
$q$- and $(l,m)$-dependence (where $l=l'$) factorizes, i.e.:
\begin{equation}
S_{ll}^{(s)m}(q,t) \cong C_l^{(s)}(t) \; \; F_s(q,t)
\label{eq12}
\end{equation}
with $C_l^{(s)}(t)$ the self part of Eq.~(\ref{eq2}) and $F_s(q,t)
\equiv S_{00}^{(s)0}(q,t)$, the self part of Eq.~(\ref{eq9}). The reader
should note, that Eq.~(\ref{eq12}) is assumed to hold for all $m$, and
that the factorization is trivial for $q=0$. To check the validity of
Eq.~(\ref{eq12}) for $q>0$, we show $S_{ll}^{(s)m}(q,t)$ and
$C_l^{(s)}(t) \cdot F_s(q,t)$ in Fig.~\ref{fig6}  ($l=1$) and Fig.~\ref{fig7}
($l=2$) for three different $q$-values and $T=0.477$. Although the
factorization becomes worse with increasing $q$, it is still a
reasonable approximation, even for $q=10.6$. Furthermore, the quality
of the factorization is better in the $\beta$-relaxation than in the
$\alpha$-relaxation regime (at least for $l=2$), and it also becomes
better with increasing temperature.

This approximate factorization does not necessarily mean that the
coupling between the TDOF and ODOF is very weak.  The comparison of
$C_l^{(s)}(t)$ with $F_s(q,t)$ in Fig.~\ref{fig6} and Fig.~\ref{fig7}
reveals the reason why $S_{ll}^{(s)m}(q,t)$ can be approximately
factorized. For instance, $C_1^{(s)}(t)$ has decayed to 0.1 for 
$t \cong 2 \cdot 10^4$, whereas at this time the value of $F_s(q=2.8,t)$ 
is still around 0.85, i.e.  the
ODOF relax much faster than the TDOF. This is consistent with our
observation that in the time span of the orientational correlation
time, as deduced from $C_1^{(s)}(t)$ at the lowest temperature, the
average center of mass positions change only a fraction (about 30 $\%$)
of the mean distance between the molecular centers.  We stress
that this is different to the MD-simulation of supercooled water.
There, $F_s(q,t)$ and $C_l(t)$ relax on approximately the same time
scale \cite{scio1}.

We now turn to the test of the various MCT-predictions (see KKSI). We
find~\cite{kammerer_phd} that $S_{11}^{(s)m}(q,t)$ do not obey the
second scaling law, i.e. the time-temperature superposition
principle. 
This observation has already been made for the case $q=0$~\cite{r1}, 
which shows that this type of correlation function does not follow
the predictions of MCT. This situation is different for the
correlation function $S_{ll}^{(s)m}(q,t)$ with $l \geq 2$ for which 
the second scaling law holds reasonably well. The critical exponents (which are 
practically $q$-independent) for the divergence of the
relaxation time,  $\gamma_1^{(s)}$ and $\gamma_2^{(s)}$, is 1.8 and  2.45, 
respectively, where the latter value is fairly close 
to the one found for the TDOF, $F_s(q,t)$, which was
2.56~\cite{r2}. 
The exceptional role for the correlators with $l=1$ is due to the 
existence of $180^{\circ}$-jumps of the molecular axis~\cite{r1}, 
since the  Legendre polynomial $P_1(\cos\theta)$ is sensitive on 
reorientations by $180^{\circ}$. The same is true for all $P_l(\cos\theta)$
with $l$ odd. But the weight of $P_l(\cos\theta)$ for $\theta \approx 0^{\circ}$
and $\theta \approx 180^{\circ}$ decreases with increasing $l$.
Since the second scaling law holds for $l \neq 1$, we can
restrict ourselves in the following to the analysis of the 
correlation functions at the {\it lowest} temperature.

In Fig.~\ref{fig9} we investigate the validity of the
first scaling law [Eq.~(I-4)]. This is done for $q=0$ by fitting
$C_l^{(s)}(t)$ with the critical correlator $G(t)$.
We remind the reader that this fit is performed for {\it fixed} values 
$\lambda= 0.76$ and $t_{\sigma} = 69$ as obtained from the similar fit of 
$F(q_{max},t)$. More details on this analysis can be found in 
section~\ref{sec4} of the preceding paper \cite{r2}.
For $l \ge 2$ (Fig.~\ref{fig9}) the critical correlator fits the data very
well over about two decades in time.  This range, however, becomes
smaller with increasing $l$, which may indicate that corrections to the
asymptotic law become more important for large $l$. If one uses
$\lambda$ and $t_{\sigma}$ (cf. (I-4)) as free fit parameters, the
resulting fits follow the data longer by additional one to two orders
of magnitude in time.  (We note that even $C_1^{(s)}(t)$ can be fitted
reasonably well with $G(t)$. Since we have shown in Ref.~\cite{r1} that
for this correlation function the first scaling law does not hold, one
might argue that it does not make sense to analyze $C_1^{(s)}$ in the
way proposed by MCT. However, we find that the violation of the second
scaling law is only weak and therefore it is not unreasonable to make
such an analysis.) The
so obtained values for $\lambda$ increase towards one with increasing
$l$ and reach, e.g., 0.97 for $l=6$.  We also mention that we do not
observe a critical law, Eq.~(I-6), the reason for which is likely the
strong influence of the microscopic dynamics on the early
$\beta$-relaxation regime.

We have found that these results do not change significantly for
$S_{ll}^{(s)m}(q,t)$ if $q>0$. From the fit with von Schweidler law
plus corrections, Eq.~(I-9), (not shown in Fig.~\ref{fig9}) one can
deduce the critical nonergodicity parameter $f_{ll}^{(s,c)m}(q)$, the
critical amplitude $\tilde{h}_{ll}^{(s)m}(q)$ and the correction
$\tilde{h}_{ll}^{(s,2)m}(q)$ which are shown in Fig.~\ref{fig11} for
$l$ = 1, 2 and 6, for the case $m=0$ (see KKSI for the difference
between ($h(q)$, $h^{(2)}(q)$) and ($\tilde{h}(q)$, $\tilde{h}^{(2)}(q)$)). 
We note that the result for $l=1$ was obtained for $\lambda = 0.76$ and 
a shift of the time scale to $t_{\sigma}' = 10$.
Due to the approximate
factorization property, the $q$-dependence of $f_{ll}^{(s,c)m}(q)$ is
given by that of $f^{(s,c)}(q) \equiv f_{00}^{(s,c)0}(q)$. The
functions $f_{ll}^{(s,c)m}(q)$ decrease with increasing $l$, as
expected from Fig.~\ref{fig9}. The variation of the critical amplitude
$\tilde{h}_{ll}^{(s)m}(q)$ and the correction
$\tilde{h}_{ll}^{(s,2)m}(q)$  with $q$ is similar to that for $l=l'=0$
(cf. Fig.~13 of KKSI) with the exception that these quantities do not
vanish for $q \rightarrow 0$.

The $\alpha$-, $\beta$- and the microscopic time scale can be better
visualized from the imaginary part $(\chi^{(s)''})_{ll}^m(q,\omega)$ of
the dynamical susceptibility as a function of $\omega$, which is shown
for $m=0$ in Fig.~\ref{fig12} for $q=q_{max}$, $l=0$ and $q=0$,
$l=1,2$. The microscopic peak is at about $\omega = 1$ for all these
values of $l$. Whereas the position of the $\alpha$-peak and the
location of the minimum (for low temperatures) are approximately the
same for $l=0$ and $l=2$, these positions are shifted to higher
frequencies by about one decade for $l=1$. 
We believe that this shift relates to the $180^{\circ}$-jumps of the 
molecules (see Ref.~\cite{r1}), because these jumps do not affect the 
correlators with even $l$, but those with odd value of $l$, and 
particularly those with $l=1$.

The rest of this section is devoted to the discussion of the collective
correlators $S_{ll}^{m}(q,t)$, which are presented in Fig.~\ref{fig13}
for $q=2.8$ (the position of the main peak of $S_{11}^{0}(q)$ (cf.
Fig.~\ref{fig2})) and in Fig.~\ref{fig14} for $q=6.5$ (the location of
the main peak of $S(q)=S_{00}^0(q)$ (cf. Fig.~\ref{fig2})).  Note,
that, due to symmetry (cf. section~\ref{sec2}), there are only two and
three independent correlators for $l=1$ and $l=2$, respectively.  These
correlators exhibit a strong $m$-dependence, in contrast to
$S_{ll}^{(s)m}(q,t)$. The reader should also note that $S_{11}^{1}(q,t)
< S_{11}^{0}(q,t)$ for $q = 2.8$, whereas $S_{11}^{1}(q,t) >
S_{11}^{0}(q,t)$ for $q = 6.5$.  These inequalities are related to the
fact that $S_{11}^{0}(q)$ has its main peak at $q \cong 2.8$ where
$S_{11}^{1}(q)$ does not have a maximum, whereas $S_{11}^{1}(q)$ has
its main peak at $q \cong 6.5$,  where $S_{11}^{0}(q)$ is close to a
minimum.  Similar considerations hold for the $m$- and $q$-dependence
of $S_{22}^{m}(q,t)$. These observations make it obvious that a
factorization [cf. Eq.~(\ref{eq12})] does not work for the collective
correlators.

The test of the second scaling law is shown in Fig.~\ref{fig15} for
$q=2.8$, $m=0$ and $l=1,2$.  As already found for $C_l^{(s)}(t)$ and
$C_l(t)$, i.e. the correlation functions for $q=0$, this scaling law
holds for $l=2$ but not for $l=1$. We define the $\alpha$--relaxation time
$\tau_{lm,q}(T)$ as the time it takes $S_{ll}^{m}(q,\tau_{lm,q})$ to
decay to the value of $1/e$. The temperature dependence of
$\tau_{lm,q}(T)$ is shown in Fig.~\ref{fig16}.  Fixing $T_c=0.475$, the
$\alpha$--relaxation times obey a power law (I-10) over about 2 - 3
decades in time. For the corresponding exponent $\gamma$ one obtains
approximately 1.9 for $l=1$ and 2.5 for $l=2$ with no significant 
$q$-dependence. Again the $\gamma$-values for $l=2$
(and the same remains true for $l=3,\ldots,6$)
fit with that for $l=0$, which was around 2.55 (see KKSI), 
whereas the value of $\gamma$ for $l=1$ is quite different.

The test of the first scaling law by fitting the  time dependence of
$S_{ll}^m(q,t)$ with the critical correlator is done in
Fig.~\ref{fig18} for $l=2$, $m=0$. This fit (again with $\lambda=0.76$ 
and $t_{\sigma}=69$) works well for different values of $q$.
From the fit with the von Schweidler law plus
correction, Eq.~(I-9), (not shown in Fig.~\ref{fig18})
we compute the critical nonergodicity parameter $f_{ll}^{c,m}(q)$, the
critical amplitude $\tilde{h}_{ll}^{m}(q)$ and the correction
$\tilde{h}_{ll}^{(2)m}(q)$, shown in Figs.~\ref{fig19} and \ref{fig20}
for, respectively, $l=1$ and $l=2$. Although we have seen, that $l=1$ is 
rather special, we have analyzed the corresponding correlators at the 
lowest temperature and have included its result.
For reference we also show in Figs.~\ref{fig19} and \ref{fig20} 
the static correlator  $S_{ll}^{m}(q)$ and
the $\alpha$-relaxation time $\tau_{lm,q}(T)$ for $T=0.477$. These
quantities possess the same characteristic $q$-dependence already found
for the corresponding quantities of the TDOF, i.e. for $l=l'=m=m'=0$
(cf. Figs. 18 and 19 of KKSI).  This means that (i) $\tau_{lm,q}$ and
$f_{ll}^{c,m}(q)$ are in phase and $\tilde{h}_{ll}^{m}(q)$  and
$\tilde{h}_{ll}^{(2)m}(q)$ are in anti-phase with $S_{ll}^{m}(q)$ and (ii) the
correction $\tilde{h}_{ll}^{(2)m}(q)$ is smallest at that $q$ where
$S_{ll}^{m}(q)$ has its main peak.  This latter fact is well pronounced
for $(l,m) = (1,0)$ and $(l,m) = (2,0)$ and less for the others, because there
also the $q$-dependence of $S_{ll}^{m}(q)$ is less pronounced.

\section{Discussion and conclusions}
\label{sec4}

For a system of diatomic and rigid molecules interacting via
Lennard-Jones potentials we have investigated by means of a
MD-simulation the time and temperature dependence of a general class of
$\vec{q}$-, $(l,m)$- and $(l',m')$-dependent correlators. These
correlators $S_{lm,l'm'}(\vec{q},t)$ contain the TDOF and ODOF
explicitly.

The static correlators $S_{ll'}^{m}(q)$ in the $q$-frame are not
diagonal in $l$ and $l'$. Whereas those with $l+l'$ odd are smaller
than $S(q) \equiv S_{00}^{0}(q)$ by about one order of magnitude, this
is not true for $S_{02}^{0}(q)$, where $l+l'$ is even. This different
behavior results from a head-tail symmetry which is only slightly
broken for our molecules.

Our main concern has been the investigation of the time-dependent
correlators (collective and self part) and a test of the predictions of
mode coupling theory (MCT). This has been restricted to the diagonal
correlators ($l=l'$). As a by-product we have found that the $q$- and
$(l,m)$-dependence of the self-correlators $S_{ll}^{(s)m}(q,t)$
approximately factorizes, which was demonstrated for $l=1,2$ and for
$q$  up to 10.6. The reason for this factorization is based on a faster
relaxation of the ODOF, compared to that of TDOF.

Concerning the MCT predictions, we first studied the existence of a
single transition temperature $T_c$. For the $(q,l,m)$-dependent
$\alpha$- relaxation times $\tau_{lm,q}(T)$ we have found that they can
be fitted with a power law (I-10) with $T_c = 0.475 \pm 0.01$. Thus
from the numerous correlators we have investigated, one unique
temperature $T_c$ can be located, at which the dynamics of TDOF and
ODOF crosses over from an ergodic to a quasi-nonergodic
behavior. This temperature also agrees with that obtained from the
translational diffusion constant $D(T)$.  This indicates that the TDOF
and the ODOF are strongly coupled. Values for $\gamma$ and the corresponding 
exponent parameter $\lambda$ are given in Table I for the translational 
diffusion constant and a selection of correlators. From this Table we 
observe that $\gamma$ is non-universal.
Nevertheless there seems to be
some systematic behavior.  The $\gamma$-values for all the correlators
with $l \neq 1$ correspond to $\lambda = 0.76 \pm 0.03$ and are 
essentially independent of $q$ and
independent of whether the collective or self correlator is
considered. A deviation from this value occurs for $\gamma_D$, the
exponent for the diffusion constant, and even a stronger one for all
correlators with $l=1$. A similar discrepancy between $\gamma_D$ and
the exponent for the $l=0$ relaxation time has been reported
before~\cite{r18}, which shows that this prediction of MCT seems to be
problematic.

This exceptional role of the $(l=1)$-correlators is also observed for
the first and second scaling law of ideal MCT. A consistent picture
within ideal MCT emerges for all $q,l,m$ with $l\neq 1$.  The situation
is illustrated in Fig.~\ref{fig21} for an exponent
parameter $\lambda=0.76$.  There we  plot $(S_{ll}^m(q,t)-f_{ll}^{c,m}(q)) 
/ \tilde{h}_{ll}^m(q)$ versus $t$, which should equal in the first 
scaling regime the critical correlator $G(t)$.  All the correlators
shown follow the ``universal'' time-dependence of the critical correlator
$G(t)$ for $\lambda=0.76$. Such a behavior was also found by
Wahnstr\"om and Lewis \cite{r13} for a simple model for
orthoterphenyl.  The time range for which the correlators can be fitted
by $G(t)$ depends on $q$, $l$ and $m$ and varies between one and a half
decade (for $C_2^{(s)}(t) \equiv S_{22}^{(s)m}(0,t)$) and three decades
(for $F(q_{max},t) \equiv S_{00}^0(q_{max},t)$ ). Although this time
range increases significantly by taking $\lambda$ and the $\beta$-relaxation 
time scale $t_\sigma$ as
free parameters which seems to yield $\lambda \rightarrow 1$ for 
$l \rightarrow \infty$, we believe that the different time ranges relate to the
$(q,l,m)$-dependence of the size of the asymptotic regime.  This has
been demonstrated earlier for the TDOF of supercooled water
\cite{scio2} and for the TDOF for our molecular system in KKSI. That
the asymptotic regime depends on $q$ has recently been shown by the
analytical calculation of the next order corrections for a system of
hard spheres \cite{r5}. We also find that for the correlators with $l =
l' \ge 0$ (with exception of $l=l'=1$) the asymptotic regime is
largest for $q_{max}^{(l)}$, the main peak of the static correlator
$S_{ll}^{m}(q)$. This is in variance with the result for
water~\cite{scio_unp}. There it has been found that the corrections are
smallest for $q=q_{FSDP}$, where $q_{FSDP}$ is the position of the
first sharp diffraction peak and not that of the main peak of $S(q)$.
This difference probably relates to the different types of glass forming
liquids. Water is a network former due to covalent bonding mechanism,
which is absent for our model liquid.  The role of this correction 
to the asymptotic laws is
also supported by the fact that the $(q,l,m)$-dependence of the
critical nonergodicity parameters, shown in Fig.~\ref{fig20}, 
is only consistent with that of $f_{ll}^{c,m}(q)$ obtained
from the molecular MCT \cite{r14} for the present liquid of diatomic
molecules, if the next order correction to the von Schweidler law (cf.
Eq.~(I-9)) is taken into account.

The result shown in Fig.~\ref{fig21} also demonstrates the validity 
of the factorization of $(q,l,m)$- and $t$-dependence of the various
correlators on the time scale of $t_\sigma$. For simple liquids, i.e. 
for $l=m=0$, this is a prediction of MCT~\cite{mct,gotze91}.
There it has been shown that the vertices of the mode coupling 
terms are positive for a simple, one-component liquid, which, however, 
is not true anymore for molecular liquids \cite{r7}. Since the 
factorization theorem only requires that the largest eigenvalue of 
a certain stability matrix (see Ref.~\cite{gotze91}) is non-degenerate, 
for which the positivity of the vertices is sufficient but not necessary, 
we still believe that this non-degeneracy is generic and that
therefore the factorization theorem holds for molecular liquids as
well. In the case that a system exhibits a type-B
transition~\cite{gotze91}, this non-degeneracy and hence the
factorization is guaranteed. 

The exceptional behavior for the correlators with $l=1$ has also been 
observed in the susceptibility (cf. Fig.~\ref{fig12}). The position of
the minimum between $\alpha$- and microscopic peak of
$(\chi^{(s)''})_{ll}^m(q,\omega)$ is approximately the same for $l=0$
and $l=2$, but not for $l=1$. For the latter it is shifted to higher
frequencies by about one order of magnitude. It is interesting that this
result resembles the experimental results for some glass forming
liquids. For instance it has been stressed by Cummins {\it et al.}
\cite{r15} , that light scattering data which may include contributions
from both, $l=0$ and $l=2$, are consistent with the spectra obtained
from neutron scattering (which is only $l=0$), but not with those from
dielectric measurements. This is nicely demonstrated for glycerol by
Lunkenheimer {\it et.~al.} \cite{r16,r17}. The situation illustrated in
Fig.~2 of \cite{r17} is exactly what we have found in Fig.~\ref{fig12}
for our system. The reader should also note that even the relative
weight between the intensity of $\alpha$- and microscopic peaks has the
same qualitative behavior in both cases, i.e. it is significantly
larger for $l=1$ than for $l=0$ and $l=2$. A similar result has been
recently found from a MD-simulation of CKN, where the orientational
dynamics (self part) of the NO$_3^-$ ion was studied for $l=1$ and
$l=2$~\cite{lebon97}. In that paper, and also for the collective 
dynamics of dipolar hard
spheres~\cite{r6,r7}, it has been concluded that the different weights of
the $\alpha$-- and microscopic peaks relate to the different numerical
values for the critical nonergodicity parameters. For $q=0$ is  
has been argued that $f_{l+1,l+1}^{(s,c)m} < f_{ll}^{(s,c)m}$ (due to $q=0$, no
$m$-dependence exists)~\cite{lebon97}. Since $f_{ll}^{(s,c)m}(q=0)$ is the
$\alpha$-relaxation strength of the corresponding susceptibility and
$(\chi^{(s)''})_{ll}^{m}(q=0)$ fulfills a sum rule (on a logarithmic frequency
scale), it becomes obvious that the ratio between the
$\alpha$-relaxation strength and the area under the microscopic peak is
larger for $l=1$ than for $l=2$.  Whether this agreement between the
susceptibilities of glycerol and that for our diatomic molecular liquid
is merely accidental or not, is, however, not obvious. One has to keep
in mind, (i) that dielectric spectroscopy and light scattering measures 
the collective dynamics and not their self part and (ii)
glycerol has a permanent dipolar moment, in contrast to
our diatomic molecules. How far the dipolar interaction would change our
MD-results is not clear. In addition, we believe that the special role
of $l=1$ relates to the $180^\circ$-jumps of the molecules \cite{r1}.
Whether these jumps exist for glycerol also and whether they really
cause a shift of the minimum is uncertain.

To summarize, we may say that the results obtained in
Refs.~\cite{r1,r2} and in the present paper are consistent with MCT.
There is strong evidence for a single transition temperature, as it is
predicted from molecular MCT \cite{r7} and for the validity of the two
scaling laws, with exception of the correlators with $l=1$. Concerning
the second scaling regime we have found that the $\gamma$-exponent is
not universal in agreement with earlier work on binary liquids
\cite{r18}, but at variance with the MD-simulation for water
\cite{scio1,scio2}. It will be a challenge to clarify the discrepancy
for the $\gamma$-values. The critical law, which is part of the first
scaling regime could not be observed, due to a strong interference with
the microscopic dynamics.

Acknowledgements: We thank the DFG, through SFB 262, for financial
support. Part of this work was done on the computer facilities of the
Regionales Rechenzentrum Kaisers\-lautern.

\newpage
\begin{figure}[f]
\psfig{file=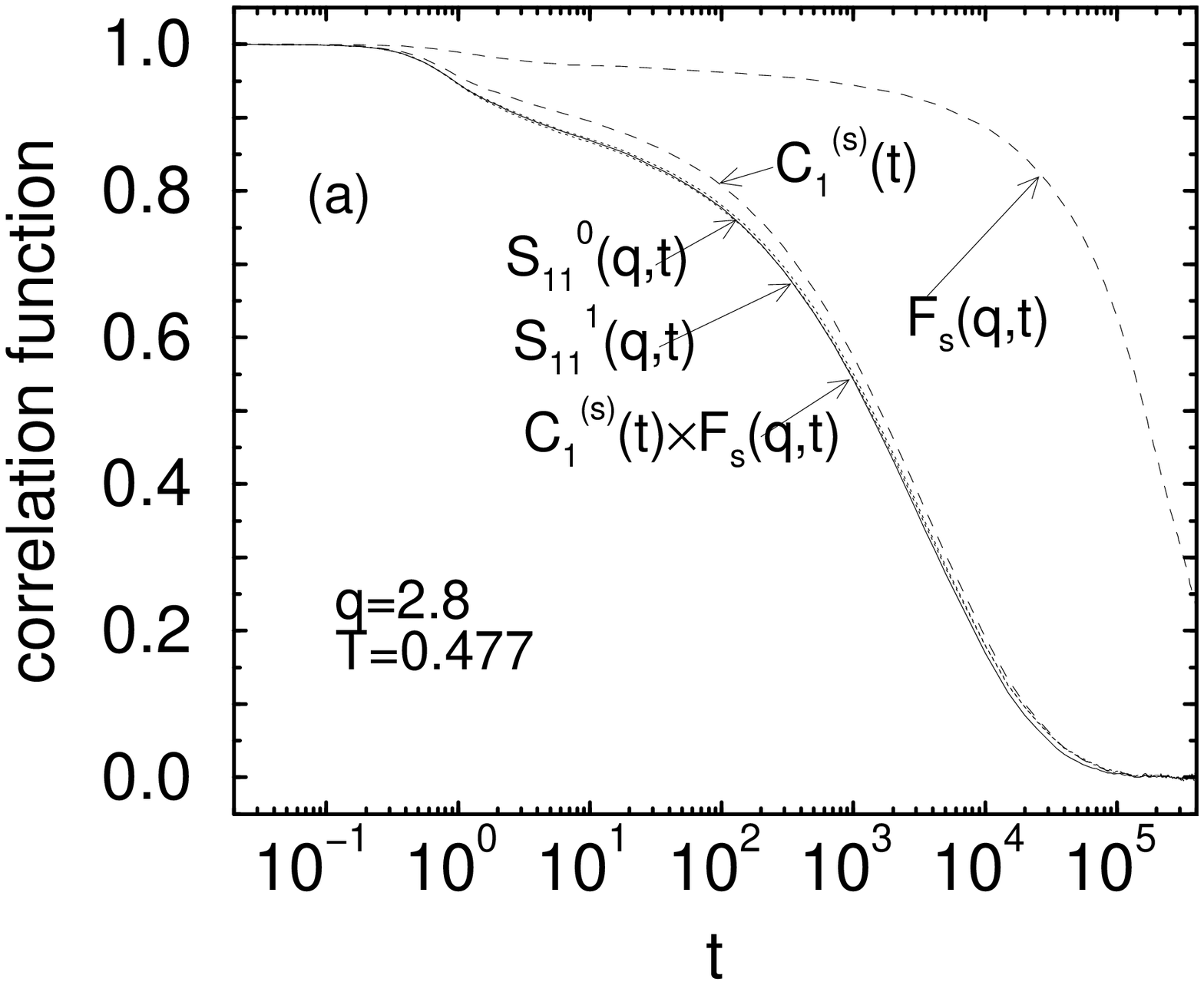,width=13cm,height=9cm}
\end{figure}
\begin{figure}[f]
\psfig{file=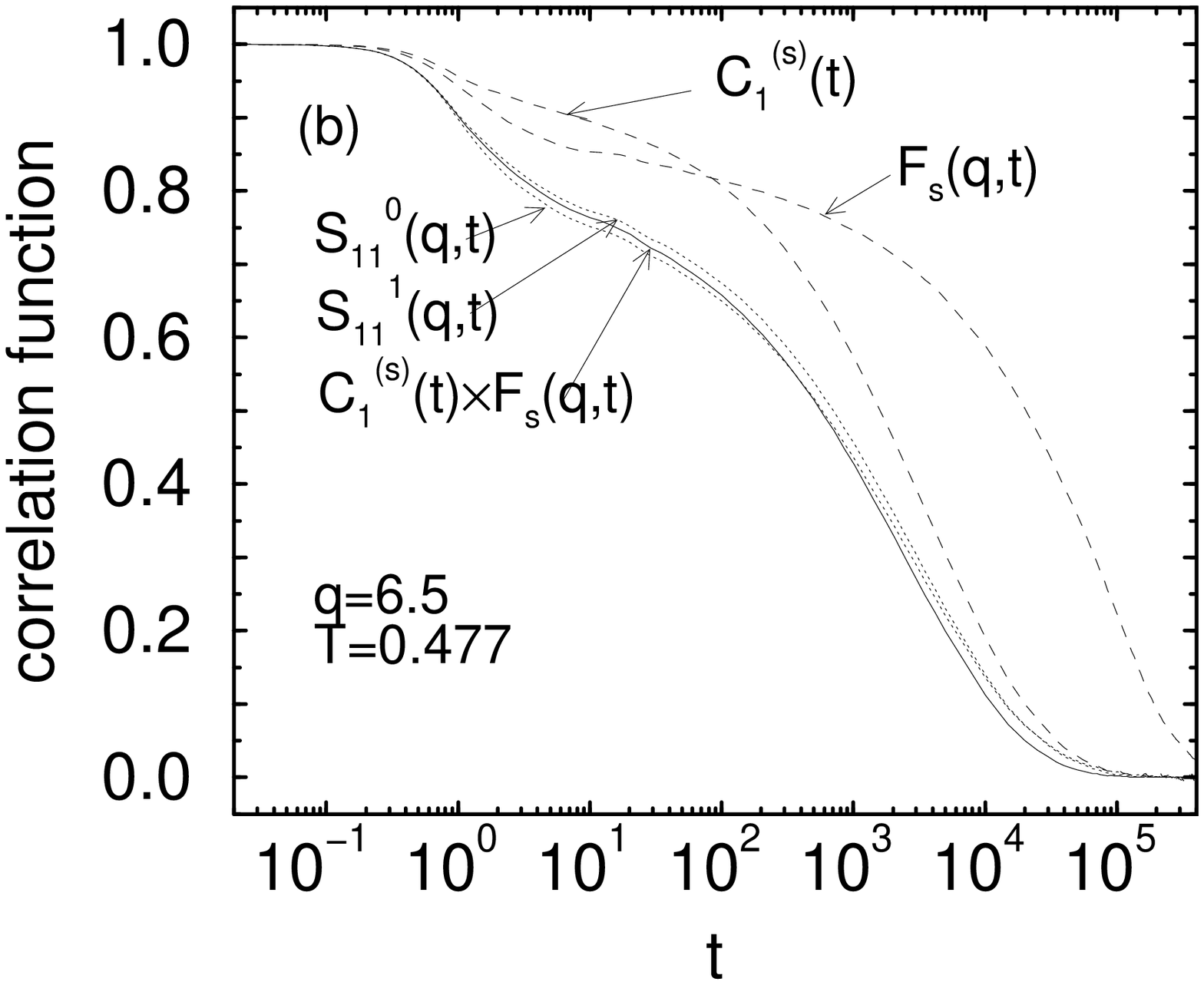,width=13cm,height=9cm}
\end{figure}
\begin{figure}[f]
\psfig{file=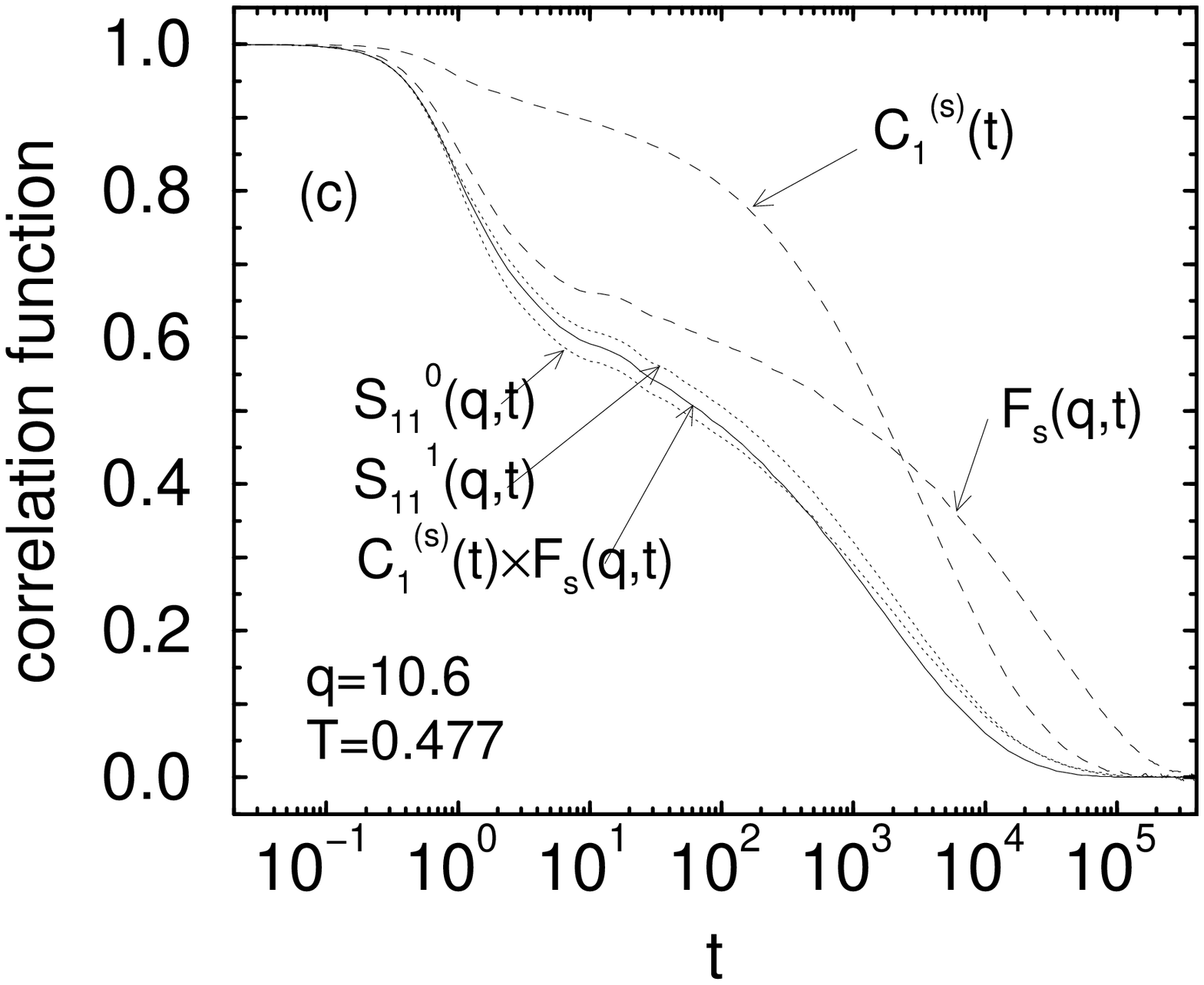,width=13cm,height=9cm}
\caption{
$S_{11}^{(s)m}(q,t)$ for $m=0,1$ (dotted lines), $C_1^{(s)}(t)$ and
$F_s(q,t)$ (dashed lines) and $C_1^{(s)}(t) \cdot F_s(q,t)$ (solid
line) versus $t$ for $T=0.477$ and (a) $q=2.8$, (b) $q=6.5$, (c)
$q=10.6$.}
\label{fig6}
\end{figure}
\begin{figure}[f]
\psfig{file=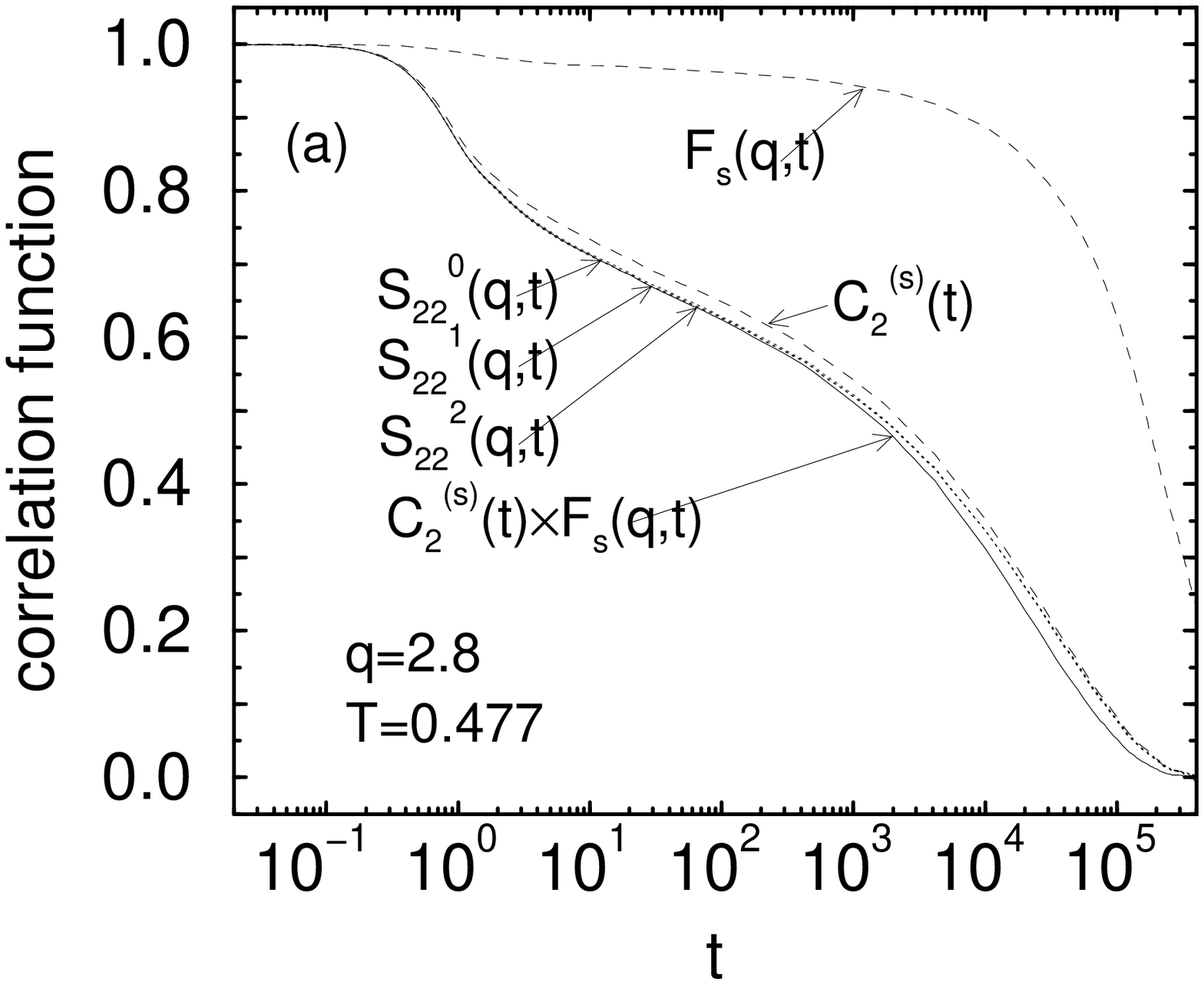,width=13cm,height=9cm}
\end{figure}
\begin{figure}[f]
\psfig{file=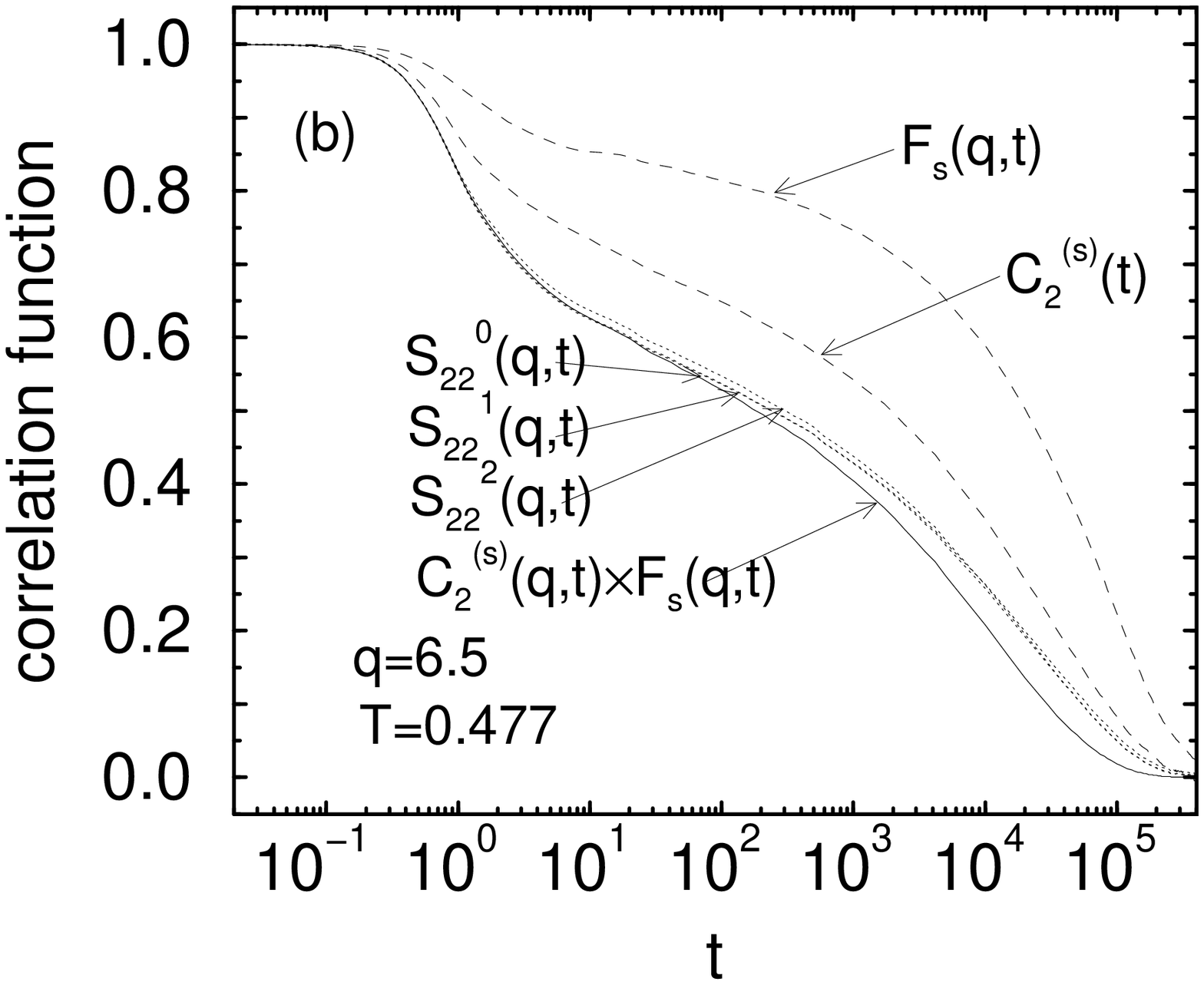,width=13cm,height=9cm}
\end{figure}
\begin{figure}[f]
\psfig{file=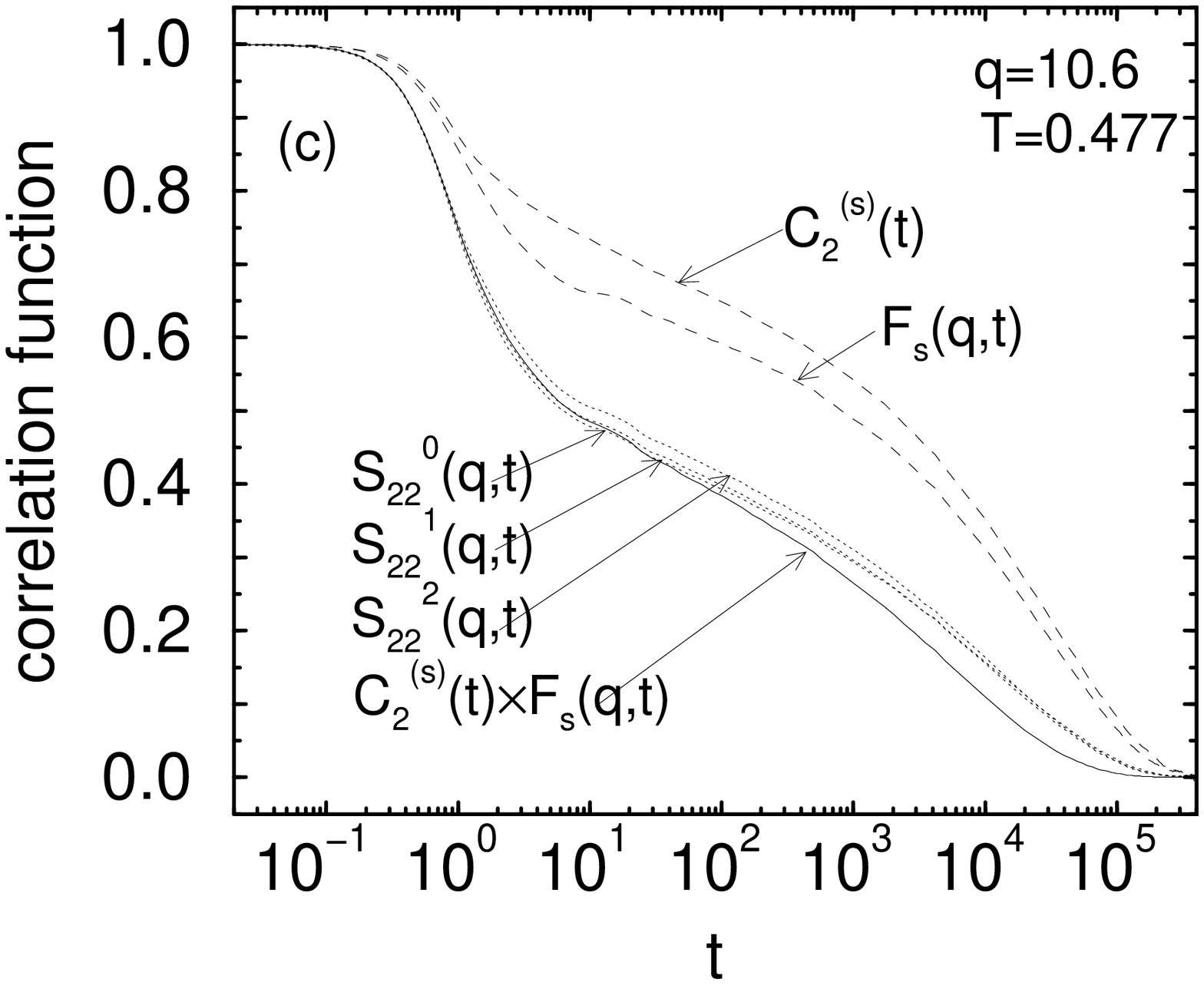,width=13cm,height=9cm}
\caption{
$S_{22}^{(s)m}(q,t)$ for $m=0,1,2$ (dotted lines), $C_2^{(s)}(t)$ and
$F_s(q,t)$ (dashed lines) and $C_2^{(s)}(t) \; F_s(q,t)$ (solid line)
versus $t$ for $T=0.477$ and (a) $q=2.8$, (b) $q=6.5$, (c) $q=10.6$.}
\label{fig7}
\end{figure}
\begin{figure}[f]
\psfig{file=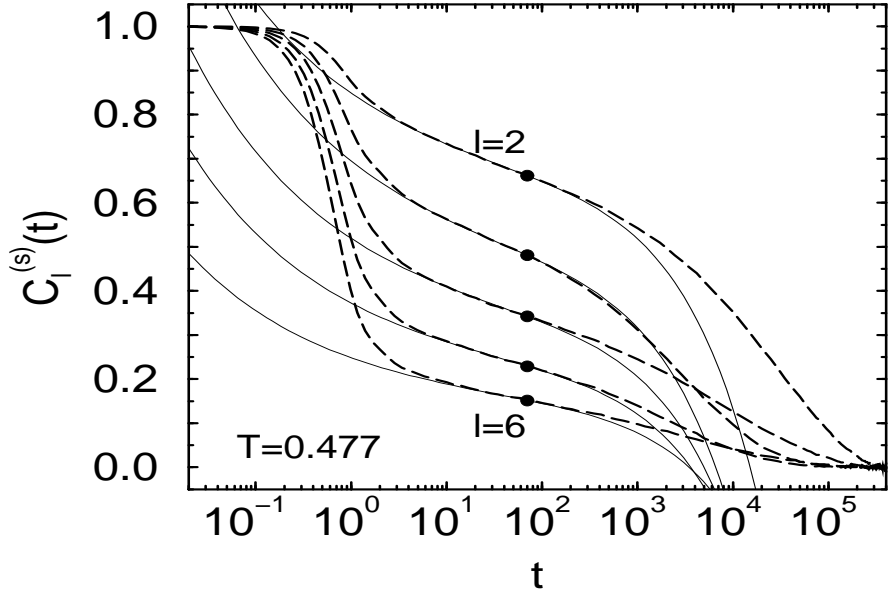,width=13cm,height=9cm}
\caption{
Time dependence of $C_l^{(s)}(t)$
for $l=2,3,\ldots,6$ (bold lines). Thin lines: $\beta$-correlator with 
$\lambda=0.76$. The circles indicate the position of the inflection 
point.}
\label{fig9}
\end{figure}
\begin{figure}[f]
\psfig{file=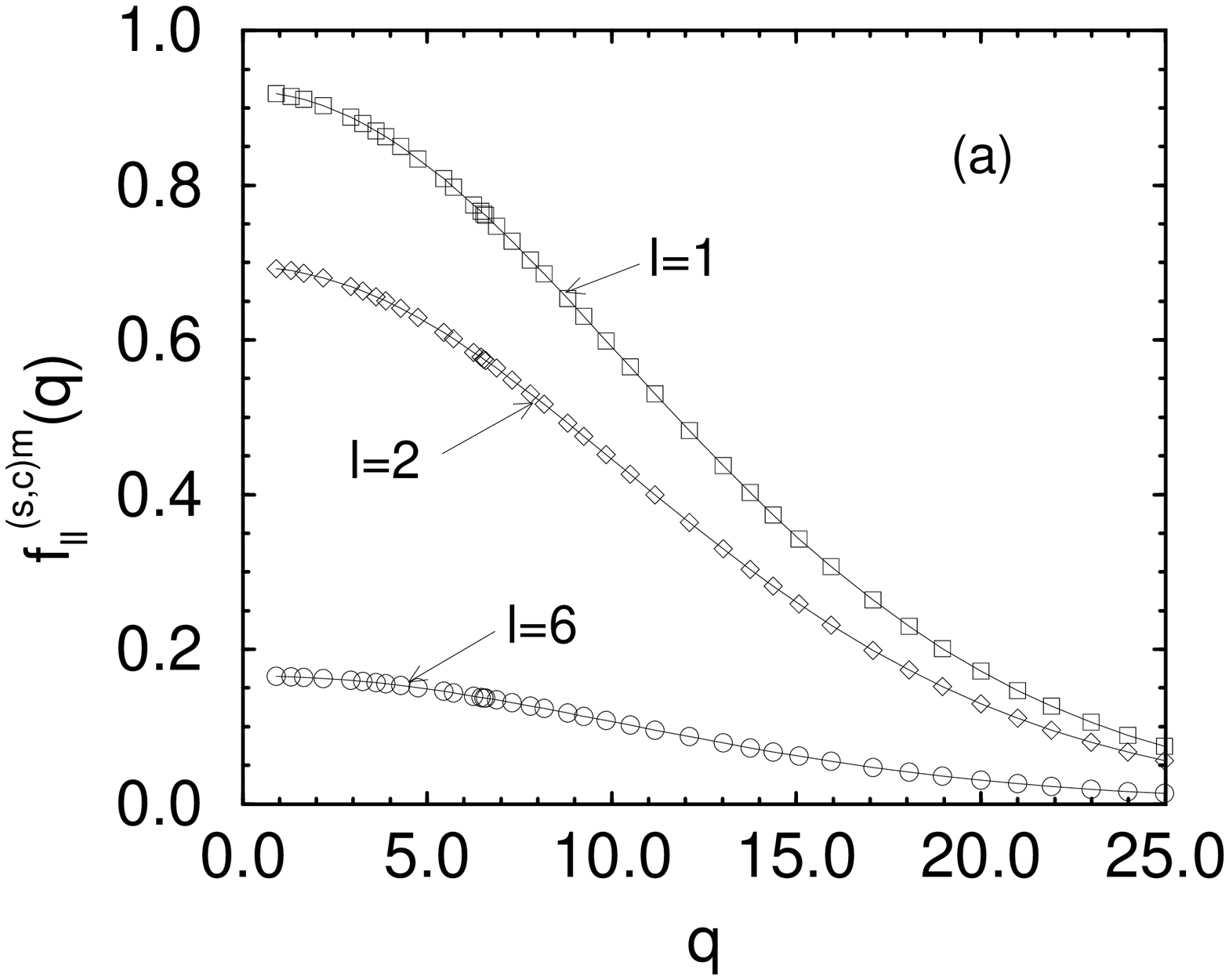,width=13cm,height=9cm}
\end{figure}
\begin{figure}[f]
\psfig{file=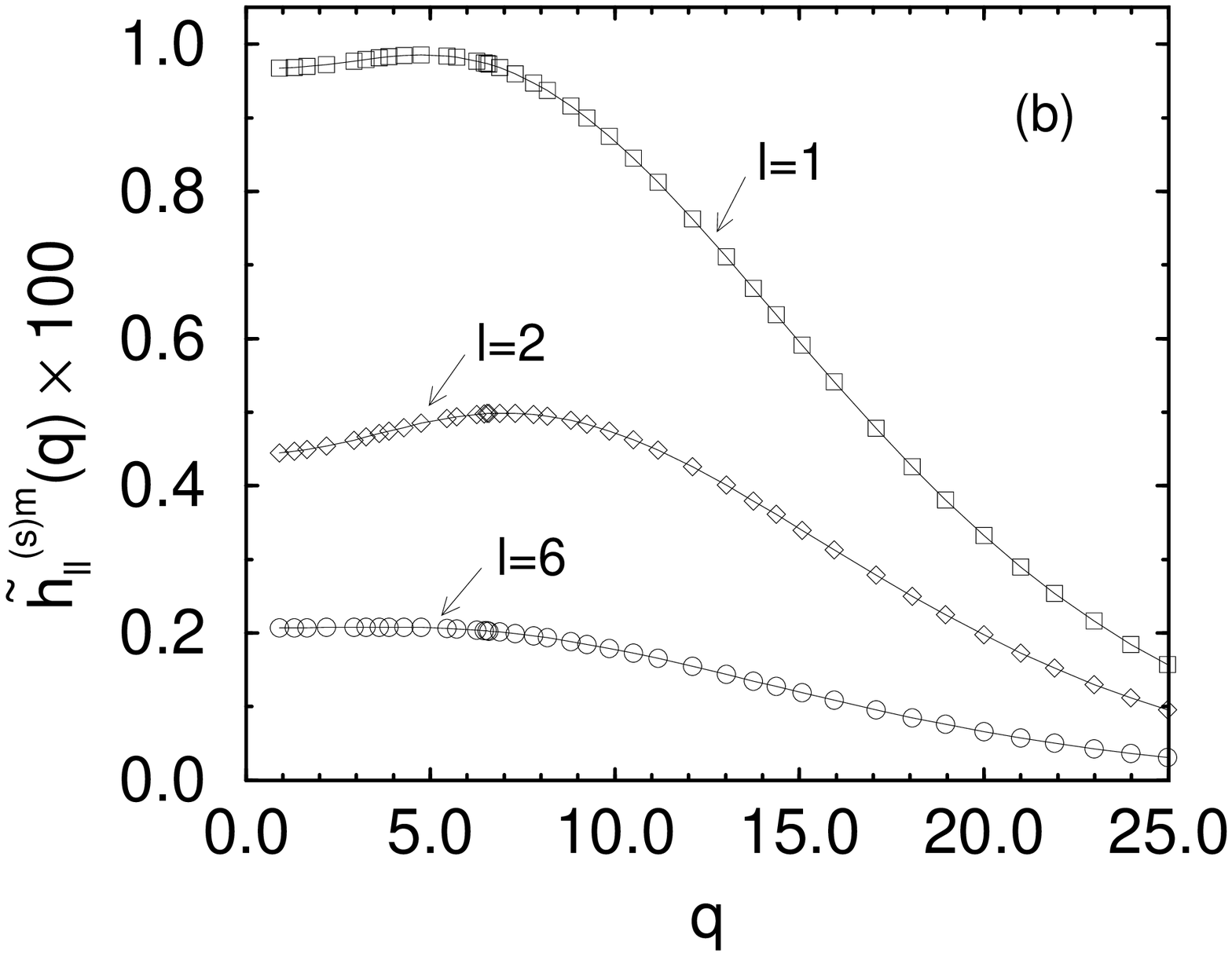,width=13cm,height=9cm}
\end{figure}
\begin{figure}[f]
\psfig{file=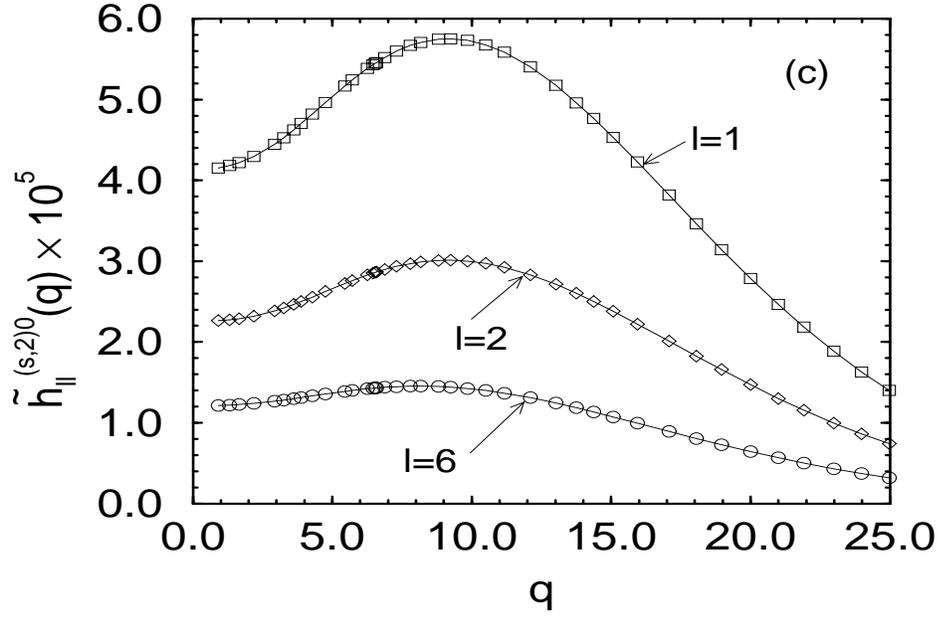,width=13cm,height=9cm}
\caption{
Wave vector dependence of $f_{ll}^{(s,c)0}$ (a), $\tilde{h}_{ll}^{(s)0}$ 
(b) and $\tilde{h}_{ll}^{(s,2)0}$ (c) for $l=2$ and $l=6$.}
\label{fig11}
\end{figure}
\begin{figure}[f]
\psfig{file=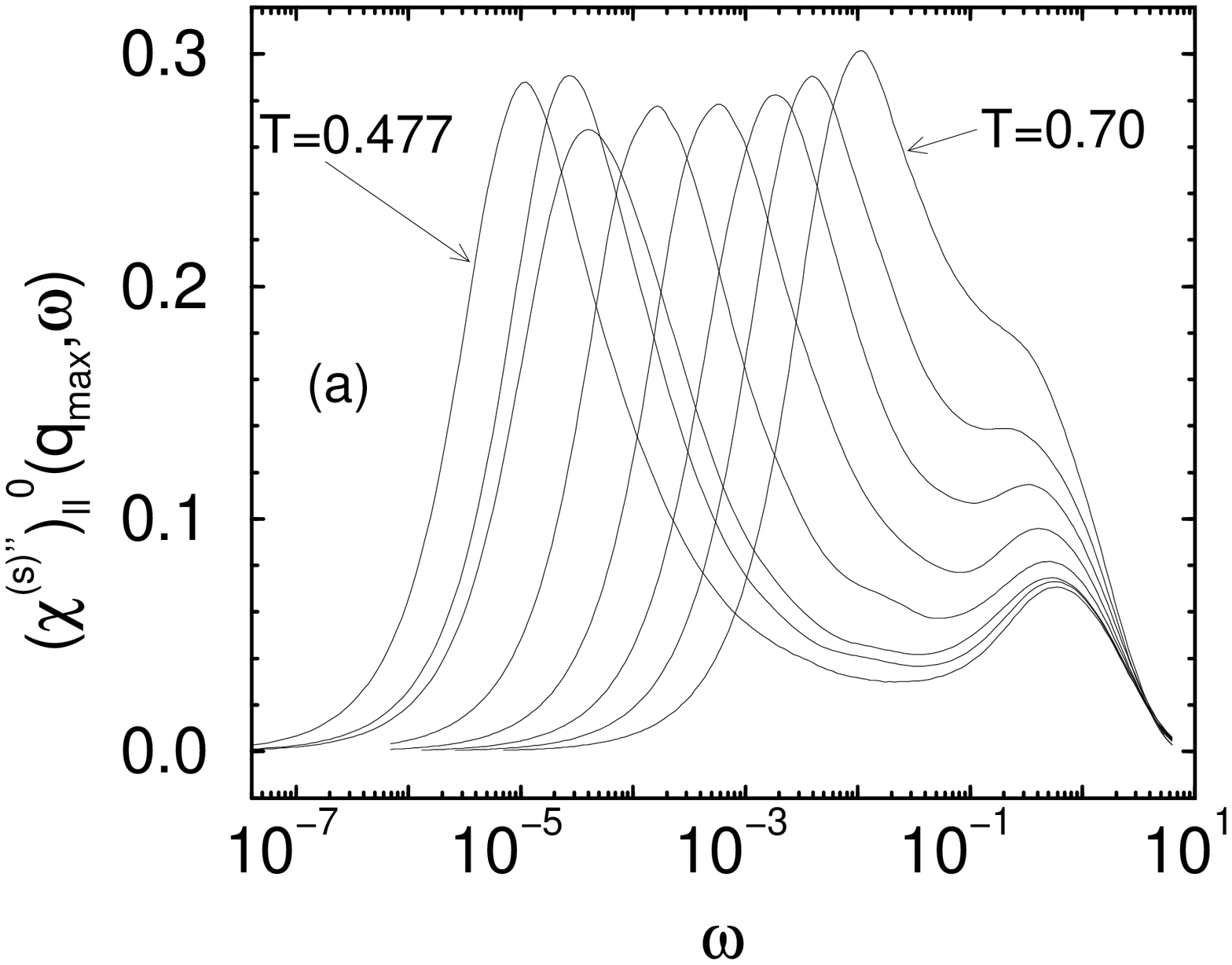,width=13cm,height=9cm}
\end{figure}
\begin{figure}[f]
\psfig{file=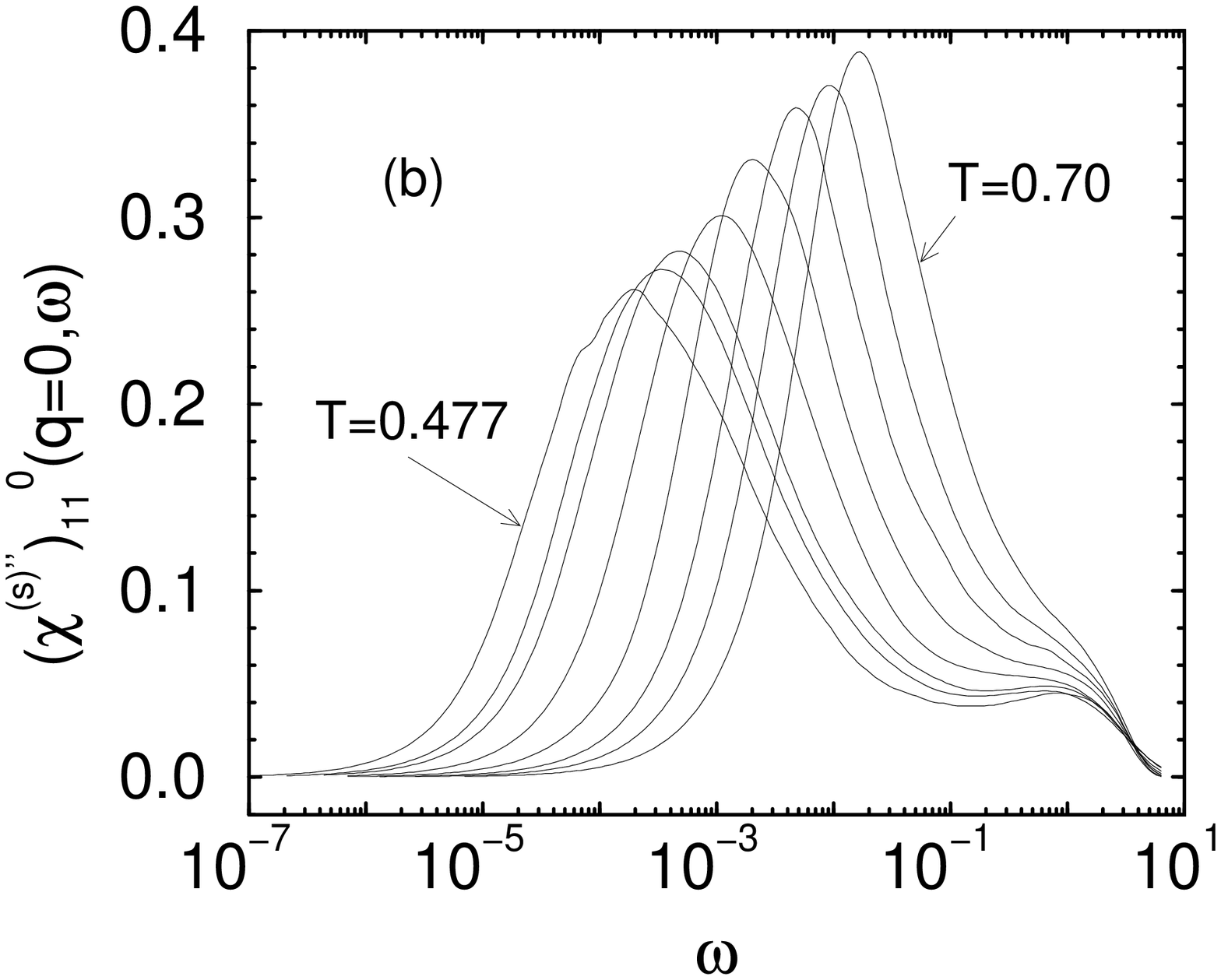,width=13cm,height=9cm}
\end{figure}
\begin{figure}[f]
\psfig{file=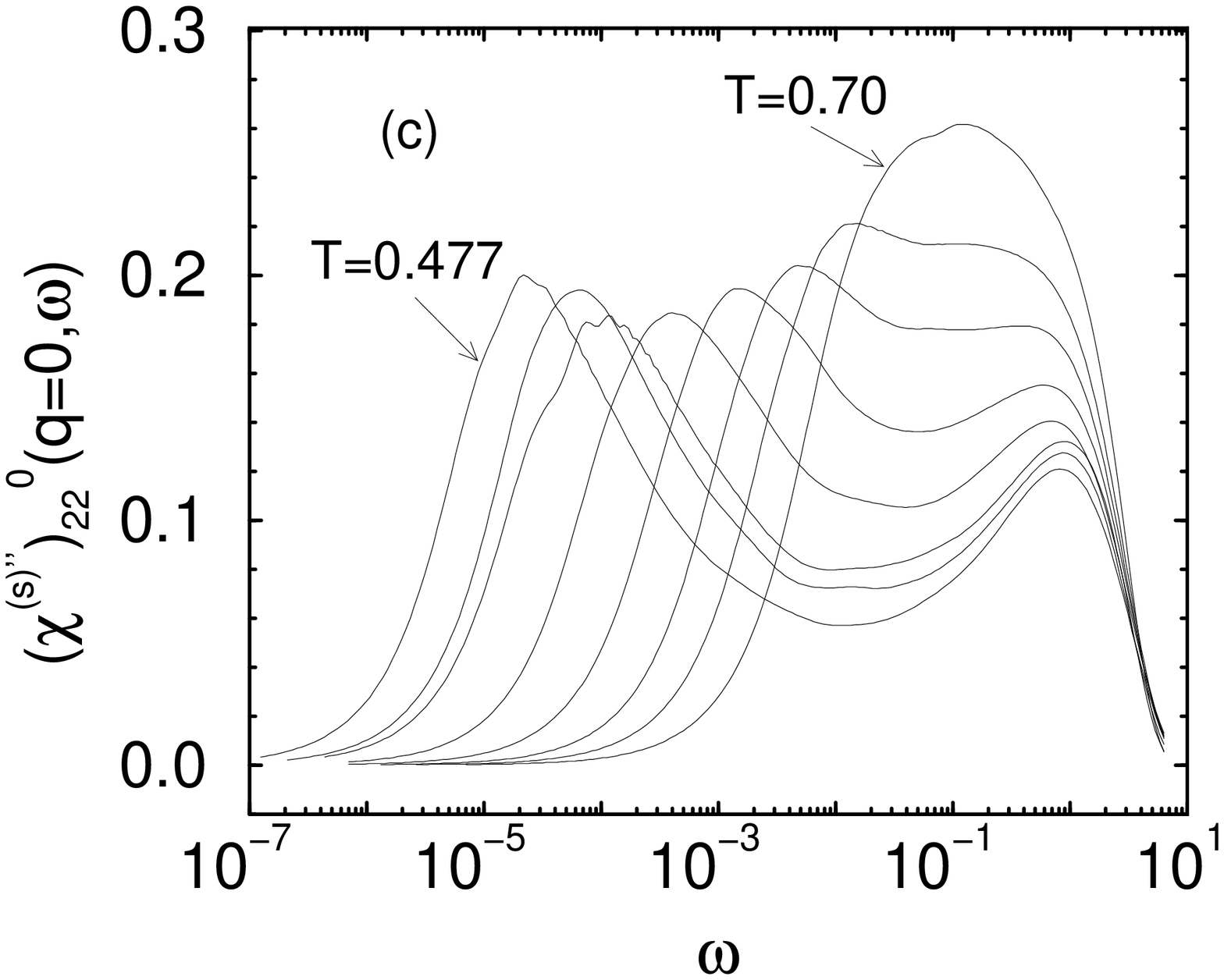,width=13cm,height=9cm}
\caption{
Imaginary part $(\chi^{(s)''})_{ll}^0(q,\omega)$ versus 
$\omega$ for the lowest investigated temperatures 
($0.477 \leq T \leq 0.7$);  
(a) $q=6.5$ and $l=0$, (b) $q=0$ and $l=1$, (c) $q=0$ and $l=2$.}
\label{fig12}
\end{figure}
\begin{figure}[f]
\psfig{file=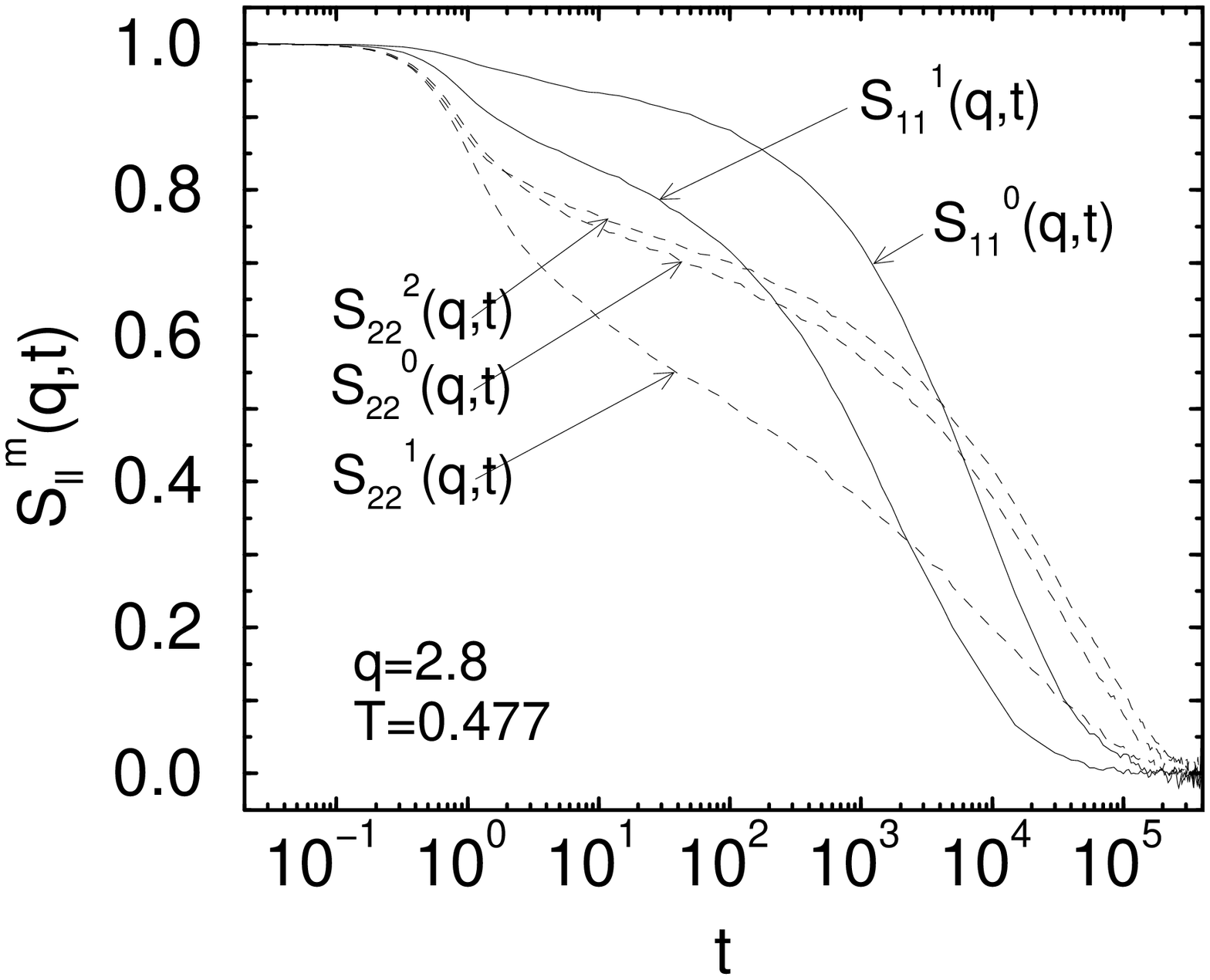,width=13cm,height=9cm}
\caption{
Time dependence of the collective correlators $S_{ll}^{m}(q,t)$  for
$q=2.8$, $T=0.477$ and $l=1$ (solid lines), $l=2$ (dashed lines).}
\label{fig13}
\end{figure}
\begin{figure}[f]
\psfig{file=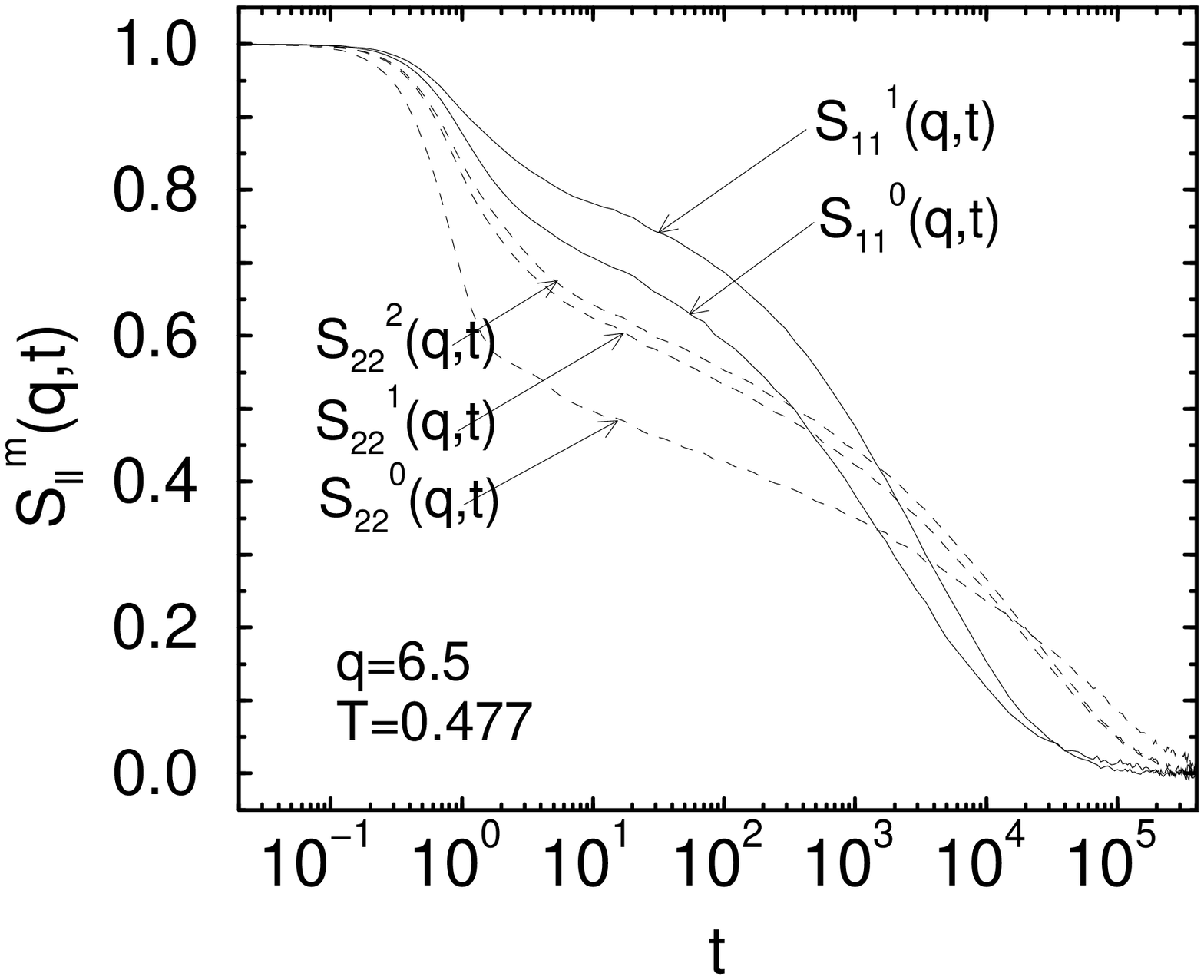,width=13cm,height=9cm}
\caption{
Time dependence of the collective correlators $S_{ll}^{m}(q,t)$  for
$q=6.5$, $T=0.477$ and $l=1$ (solid lines), $l=2$ (dashed lines).}
\label{fig14}
\end{figure}
\begin{figure}[f]
\psfig{file=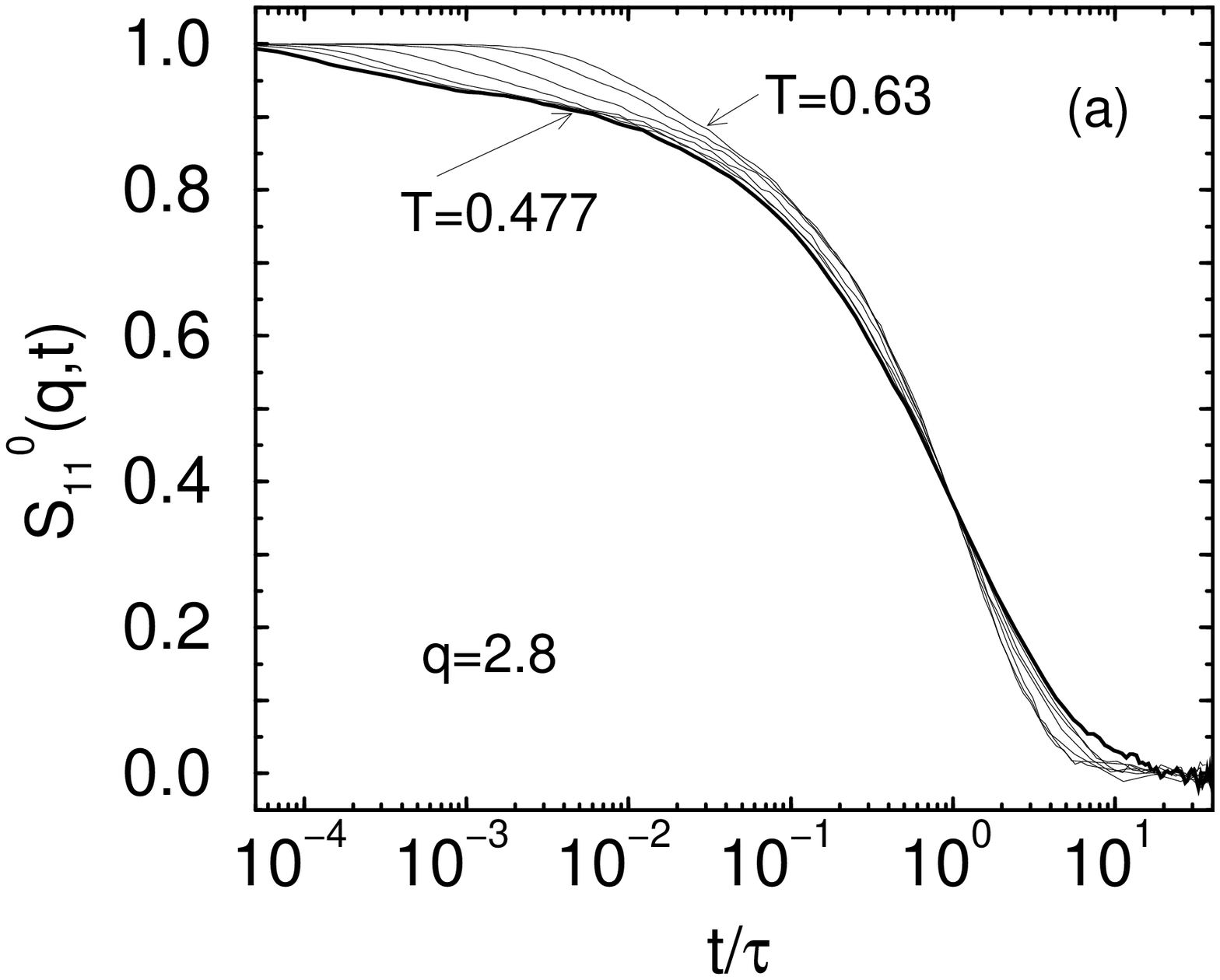,width=13cm,height=9cm}
\end{figure}
\begin{figure}[f]
\psfig{file=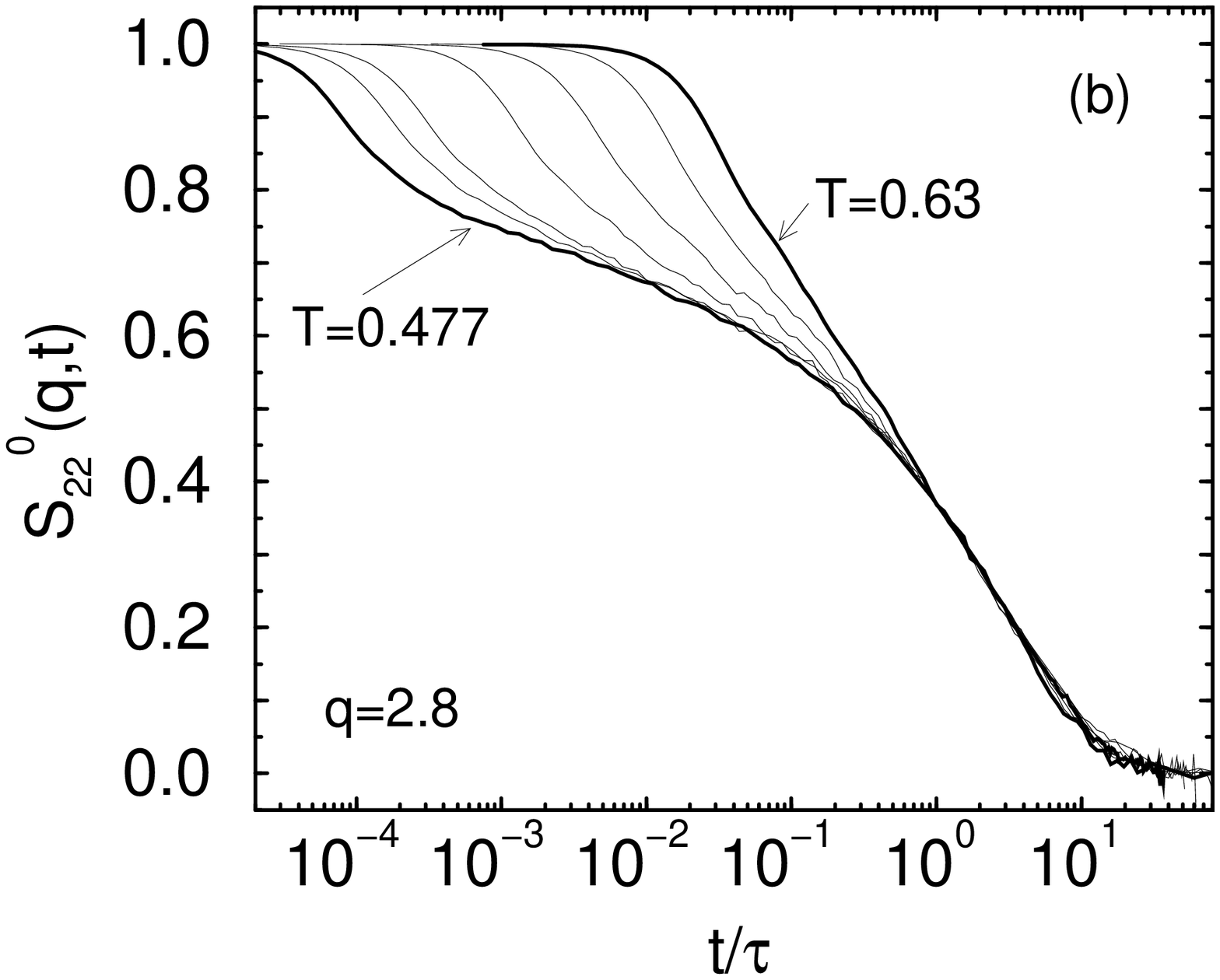,width=13cm,height=9cm}
\caption{
$S_{ll}^{m}(q,t)$ versus rescaled time for $q=2.8$, $m=0$, and 
(a) $l=1$, (b) $l=2$ for the seven lowest temperatures. The bold 
lines indicate the lowest and highest of these temperatures.}
\label{fig15}
\end{figure}
\begin{figure}[f]
\psfig{file=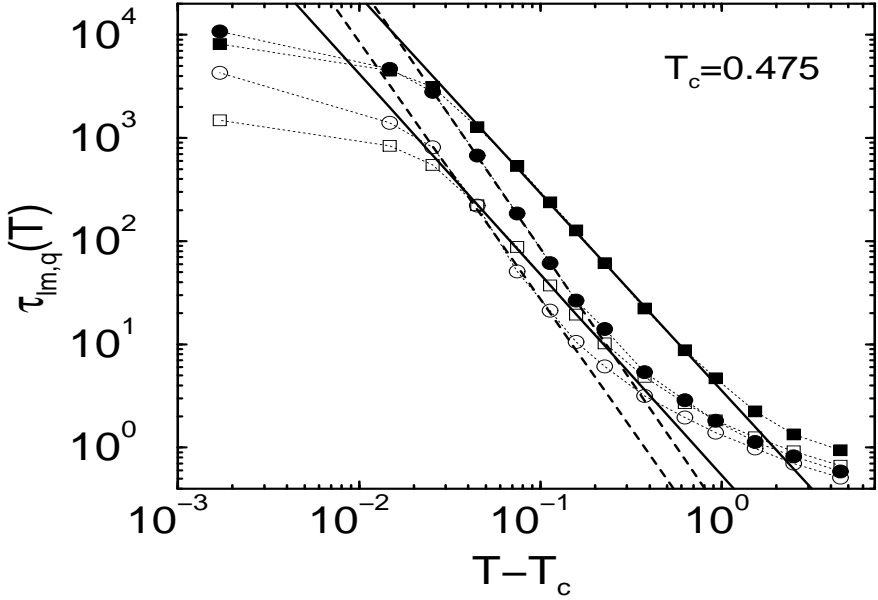,width=13cm,height=9cm}
\caption{
Relaxation times $\tau_{lm,q}(T)$ versus $T-T_c$ for $m=0$ and
$q=2.8$, $l=1$ (filled squares); $q=2.8$, $l=2$ (filled circles);
$q=7.3$, $l=1$ (open squares); $q=7.3$, $l=2$ (open circles).
$T_c=0.475$.  The bold lines represent power laws and the thin
lines are a guide to the eye.}
\label{fig16}
\end{figure}
\begin{figure}[f]
\psfig{file=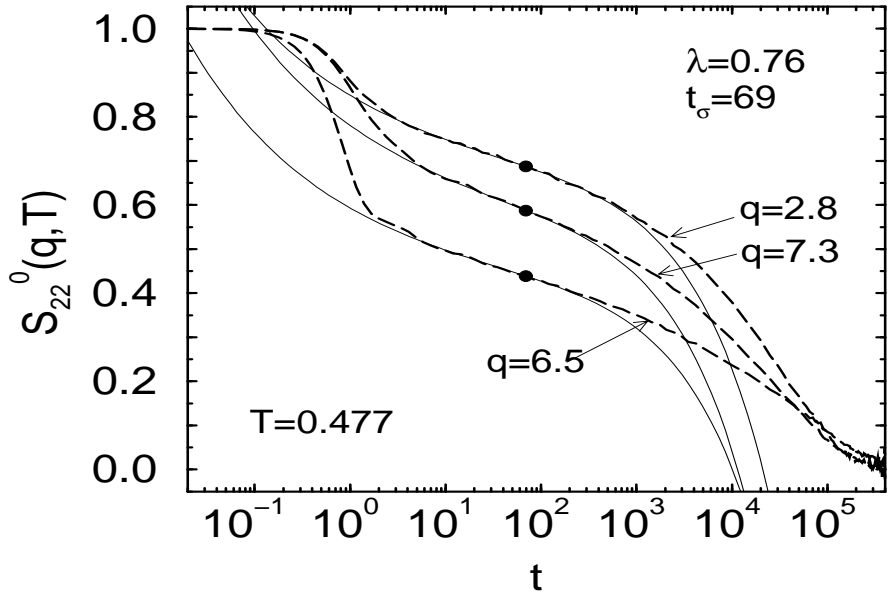,width=13cm,height=9cm}
\caption{
$S_{22}^{0}(q,t)$ (bold dashed lines) versus $t$ for $q$ = 2.8 , 6.5 
and 7.3 for the lowest temperature and the critical correlator 
(thin solid lines) with $\lambda = 0.76$ and a time scale $t_{\sigma} = 69$.
The circles indicate the position of the inflection point.}
\label{fig18}
\end{figure}
\begin{figure}[f]
\psfig{file=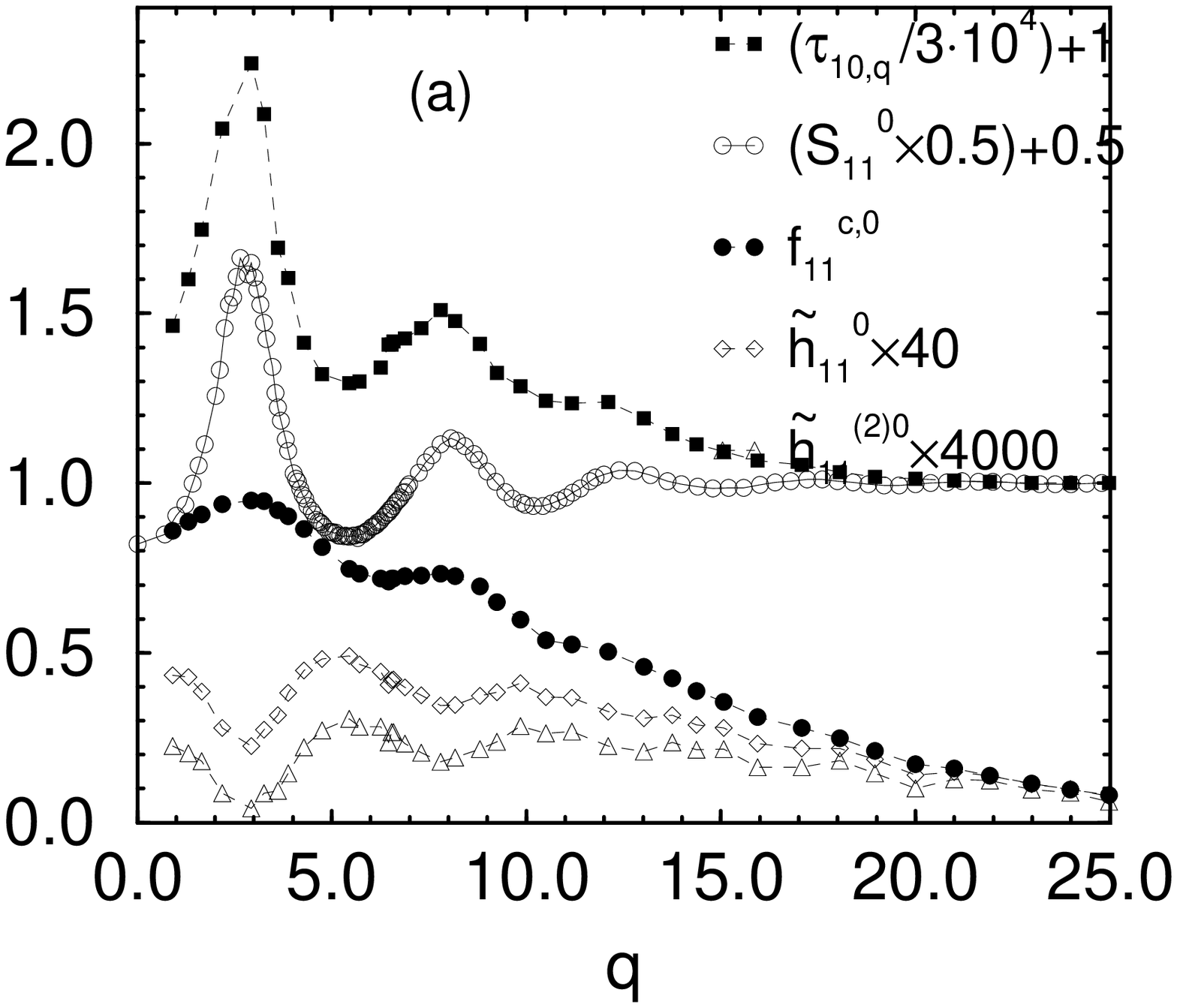,width=13cm,height=9cm}
\end{figure}
\begin{figure}[f]
\psfig{file=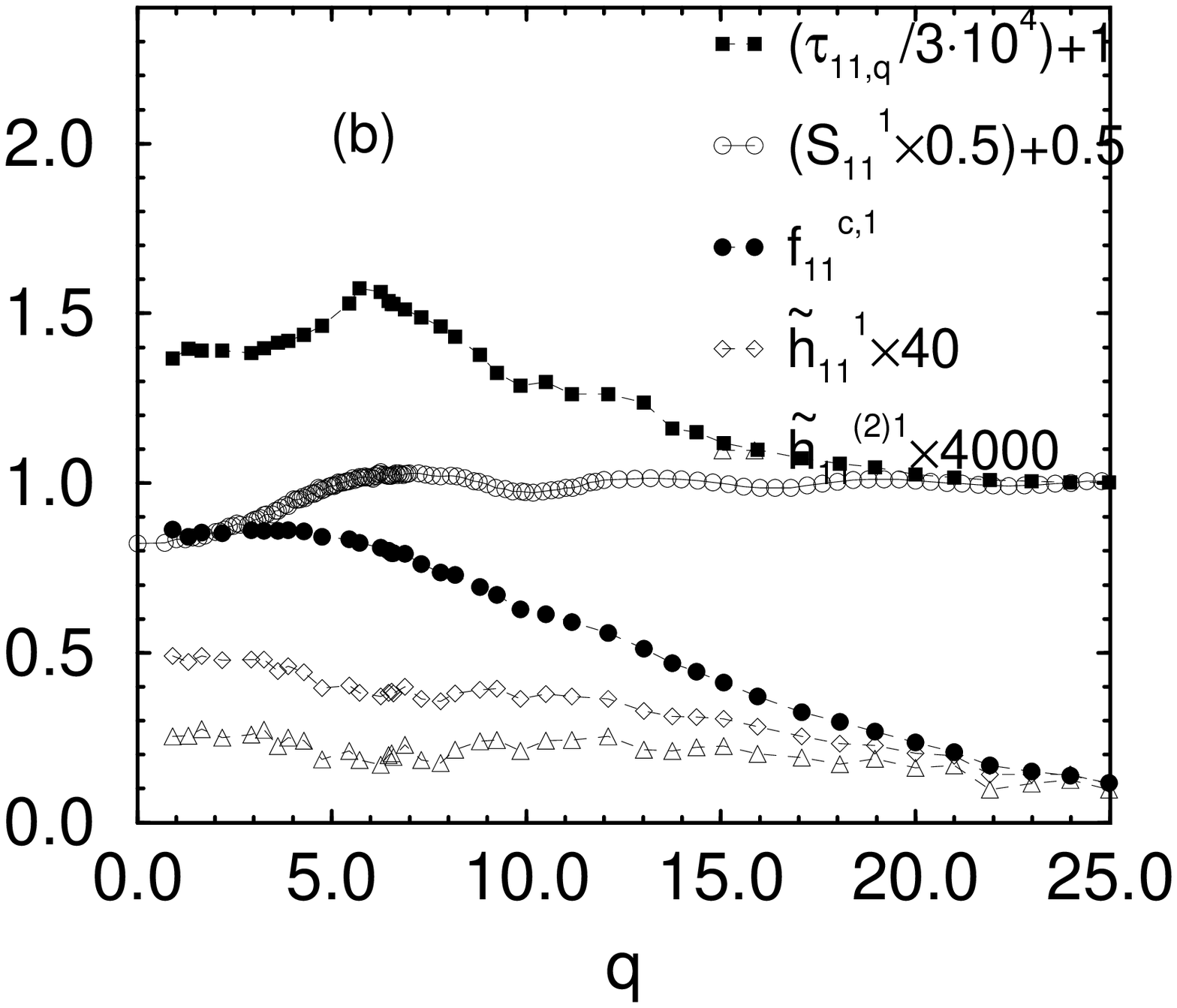,width=13cm,height=9cm}
\caption{
$f_{11}^{c,m}(q)$ (filled circles), 
$\tilde{h}_{11}^{m}(q)$ (open diamonds), 
$\tilde{h}_{11}^{(2)m}(q)$ (open triangles),
$S_{11}^{m}(q)$ (open circles)
and $\tau_{1m,q}(T)$ (filled squares) versus $q$ for $T=0.477$;
(a) $m=0$ and (b) $m=1$.}
\label{fig19}
\end{figure}
\begin{figure}[f]
\psfig{file=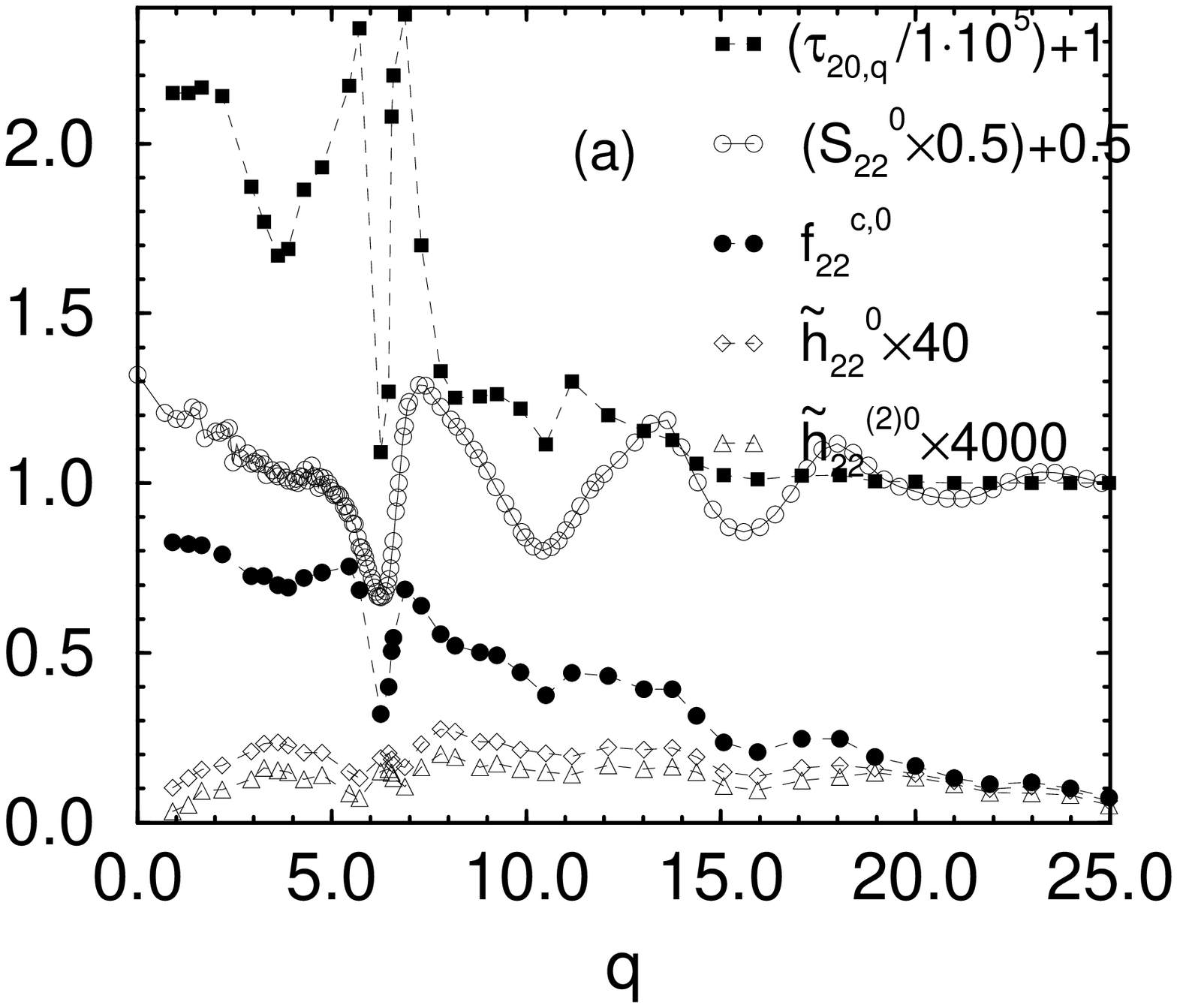,width=13cm,height=9cm}
\end{figure}
\begin{figure}[f]
\psfig{file=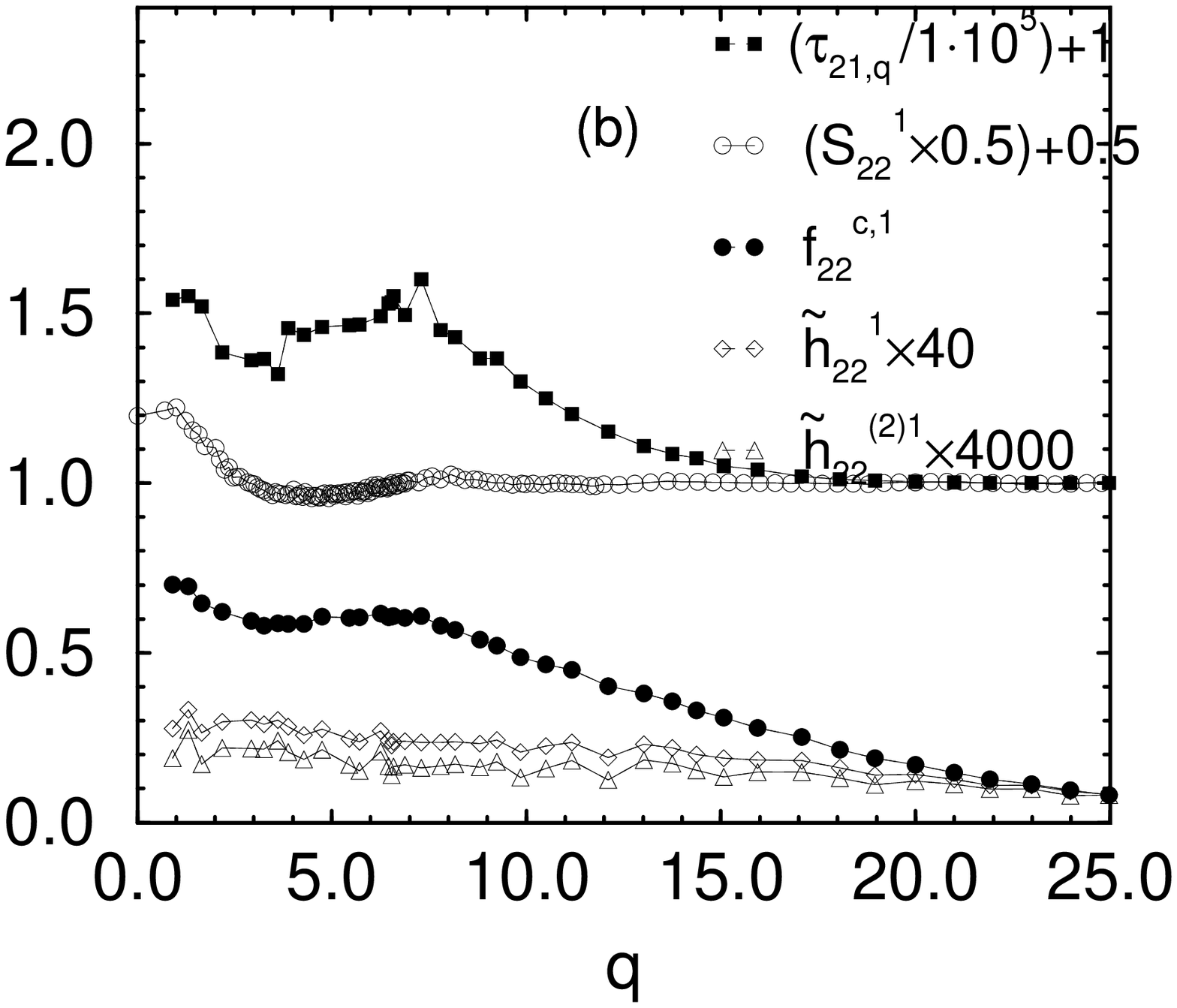,width=13cm,height=9cm}
\end{figure}
\begin{figure}[f]
\psfig{file=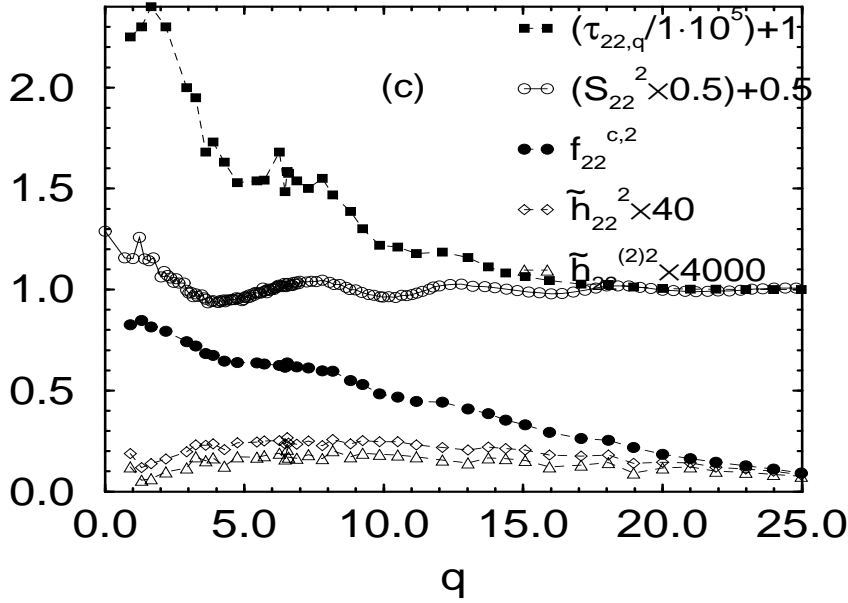,width=13cm,height=9cm}
\caption{
$f_{22}^{c,m}(q)$ (filled circles), 
$\tilde{h}_{22}^{m}(q)$ (open diamonds), 
$\tilde{h}_{22}^{(2)m}(q)$ (open triangles),
$S_{22}^{m}(q)$ (open circles)
and $\tau_{2m,q}(T)$ (filled squares) versus $q$ for $T=0.477$;
(a) $m=0$, (b) $m=1$ and (c) $m=2$.}
\label{fig20}
\end{figure}
\begin{figure}[f]
\psfig{file=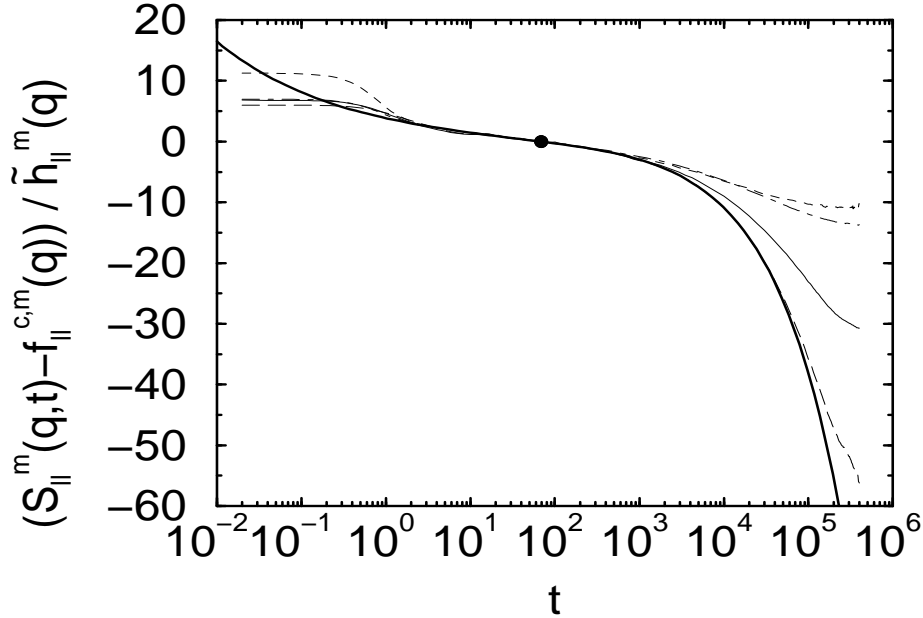,width=13cm,height=9cm}
\caption{
Time dependence of various correlators for $T=0.477$ shifted by the
corresponding nonergodicity parameter $f_{ll}^{c,m}(q)$ and
subsequently divided by the critical amplitude $\tilde{h}_{ll}^{m}(q)$.
$T=0.477$.  $F_s(q_{max},t)$ (solid line), $F(q_{max},t)$ (long dashed
line), $F(q_{min},t)$ (short dashed line), and $C_2^{(s)}(t)$
(dashed-dotted line). The bold line is the critical correlator $G(t)$
for $\lambda = 0.76$ and $t_{\sigma}=69$.}
\label{fig21}
\end{figure}
\begin{table}[tbp]
\begin{center}
\begin{tabular}{|r||r|r|r|r|r|r|r|}
\hline
 & $D$ & $C_1^{(s)}$ & $C_2^{(s)}$ & $C_6^{(s)}$ & 
		$F_s(q_{max})$ & $F(q_{min})$ & $F(q_{max})$  \\ \hline \hline
$\gamma$  & 2.20 & 1.66 & 2.42 & 2.80 & 2.56 & 2.47 & 2.57 \\ \hline
$\lambda$ & 0.67 & $<$0.5 & 0.73 & 0.79 & 0.76 & 0.74 & 0.76 \\ \hline
\end{tabular}
\end{center}
\caption{The $\gamma$-exponent and the corresponding exponent parameter 
   $\lambda$ (from (I-7) and (I-11)) for the translational diffusion 
   constant $D$ and various correlators.}
\end{table}


\begin{references}

\bibitem{r1}
S. K\"ammerer, W. Kob and R. Schilling, Phys. Rev. E {\bf 56}, xxxx (1997).
%
\bibitem{mct}
W. G\"otze and L. Sj\"ogren, Rep. Prog. Phys. {\bf 55}, 241 (1992).
%
\bibitem{gotze91}
W. G\"otze in {\it Liquids, freezing and the glass transition}, Eds. J. P.
Hansen, D. Levesque and J. Zinn-Justin (North-Holland, Amsterdam,
1991);
%
R. Schilling in {\it Disorder Effects on Relaxation Processes},
Eds. R. Richert and A. Blumen (Springer, Berlin, 1994);
%
Theme Issue on Relaxation Kinetics in Supercooled Liquids-Mode Coupling
Theory and its Experimental Tests, Ed. S. Yip. Volume {\bf 24}, No.
6-8 (1995) of {\it Transport Theory and Statistical Physics};
%
W. Kob, p. 28 in {\it Experimental and Theoretical Approaches to
Supercooled Liquids: Advances and Novel Applications}, Eds.: J.
Fourkas, D. Kivelson, U. Mohanty, and K. Nelson (ACS Books,
Washington, 1997).
%
\bibitem{r2}
S. K\"ammerer, W. Kob and R. Schilling, preceding paper.
%
\bibitem{r4}
C.~G.~Gray and K.~E.~Gubbins, {\it Theory of Molecular Liquids}, 
Vol. 1, (Clarendon Press, Oxford, 1984).
%
\bibitem{r5}
T.~Franosch, M.~Fuchs, W.~G\"otze, M.~R.~Mayr and A.~P.~Singh (preprint).
%
\bibitem{r6}
T.~Scheidsteger and R.~Schilling, 6. Int. Workshop on Disordered Systems, 
Andalo (1997).
%
\bibitem{r7}
R. Schilling and T. Scheidsteger, Phys. Rev. E {\bf 56}, xxxx (1997).
%
\bibitem{r8}
Ch. Theis, Diploma thesis, (Johannes Gutenberg-Universit\"at Mainz, 1997).
%
\bibitem{kk}
K. Kawasaki, Physica A, (in press).
%
\bibitem{r9}
R. Schmitz (unpublished).
%
\bibitem{r10}
L. J. Lewis and G. Wahnstr\"om, Phys. Rev {\bf E 50}, 3865 (1994).
%
\bibitem{r11}
B. J. Berne and R. Pecora ``Dynamic light scattering'', 
Wiley, New York, 1976.
%
\bibitem{scio1}
P. Gallo, F. Sciortino, P. Tartaglia and S.-H. Chen,
Phys. Rev. Lett. {\bf 76}, 2730 (1996);
F. Sciortino, P. Gallo, P. Tartaglia and S.-H. Chen,
Phys. Rev {\bf E 54}, 6331 (1996).
%
\bibitem{kammerer_phd}
S. K\"ammerer, PhD thesis (Johannes Gutenberg-Universit\"at Mainz, 1997).
%
\bibitem{r18}
W.~Kob and H.~C. Andersen, Phys. Rev. Lett. {\bf 73}, 1376 (1994);
W.~Kob and H.~C. Andersen, Phys. Rev. E {\bf 51}, 4626 (1995);
{\it ibid.} {\bf 52}, 4134 (1995).
%
\bibitem{r13}
G. Wahnstr\"om and L. J. Lewis, Prog. Theor. Phys. (in press).
%
\bibitem{scio2}
F. Sciortino, P. Gallo, P. Tartaglia and S. H. Chen, Phys. Rev. E
{\bf 54}, 6331 (1996).
%
\bibitem{scio_unp}
F. Sciortino, L. Fabbian, S. H. Chen, and P. Tartaglia (unpublished).
%
\bibitem{r14}
C. Theis and R. Schilling (unpublished).
%
\bibitem{r15}
H. Z. Cummins, G. Li, W. Du, R. Pick, and C. Dreyfus, Phys. Rev. E
{\bf 53}, 896 (1996).
%
\bibitem{r16}
P. Lunkenheimer, A. Pimenow, B. Schiener,  R. B\"ohmer and A.
Loidl, Europhys. Lett. {\bf 33}, 611 (1996);
P. Lunkenheimer, A. Pimenow, M. Dressel, Y. G. Gonchunov,
R. B\"ohmer and A. Loidl, Phys. Rev. Lett.  {\bf 77}, 318 (1996).
%
\bibitem{r17}
P. Lunkenheimer,  A. Pimenow, M. Dressel, Y. G. Gonchunov, U.
Schneider, B. Schiener R. B\"ohmer and A. Loidl, Proc. MRS Fall
Meeting, Boston 1996.
%
\bibitem{lebon97}
M. J. Lebon, C. Dreyfus, Y. Guissani, R. M. Pick, and H. Z. Cummins,
Z. Phys. B (in press).
%
%


\end{references}
\end{document}